\documentclass[twocolumn,amstext,amssymb,floatfix,showpacs,nofootinbib]{revtex4-1}

\usepackage{amsmath}
\usepackage{lineno} 
\usepackage{ulem}

\usepackage{todonotes}
\presetkeys{todonotes}{color=green!40, size=\footnotesize}{}
\usepackage{hyperref}

\hypersetup{colorlinks=true,citecolor=blue,citebordercolor=red,linktoc=all,linkcolor=blue}

\usepackage{setspace}

\usepackage{amssymb,amsfonts,multirow}

\usepackage{epsfig}

\usepackage{amssymb,url,bm}
\usepackage{graphicx}
\usepackage{hyperref}

\usepackage{color}

\usepackage{slashed}

\usepackage{color}
\DeclareMathOperator{\Tr}{Tr}

\long\def\del #1 \enddel { }

\usepackage{epsfig}

\usepackage{xcolor,colortbl}

\definecolor{Gray}{gray}{0.85}
\definecolor{LightGreen}{rgb}{0.88, 1, 0.88}
\definecolor{Blue}{rgb}{0,1,1}
\definecolor{Lime}{rgb}{0,1,0}
\definecolor{LightCyan}{rgb}{0.88,1,1}
\definecolor{LightRed}{rgb}{1, 0.85, 0.85}
\definecolor{Red}{rgb}{1, 0, 0}
\definecolor{LightYellow}{rgb}{1, 1, 0.85}
\definecolor{Yellow}{rgb}{1,1,0.05}
\definecolor{LightBlue}{rgb}{0.87, 0.94, 1}
\definecolor{white}{gray}{1}
\definecolor{black}{gray}{0}

\definecolor{LightGray}{gray}{0.93}

\newcolumntype{G}{>{\columncolor{LightGray}}c}

\newcolumntype{?}{!{\vrule width 1pt}}
\newcolumntype{`}{!{\vrule width 1.5pt}}

\usepackage{epsfig}

\usepackage{array,mathtools,amssymb,booktabs}
\newcolumntype{C}{>{$}c<{$}}
\AtBeginDocument{
\heavyrulewidth=.16em
\lightrulewidth=.1em
\cmidrulewidth=.03em
\belowrulesep=.4ex
\belowbottomsep=0pt
\aboverulesep=.4ex
\abovetopsep=0pt
\cmidrulesep=\doublerulesep
\cmidrulekern=.5em
\defaultaddspace=.5em
}

\usepackage{hyperref}

\newcommand{\co}[2]{\Xi_{#1}(#2)}

\newcommand{\gp}{\mathcal{G}}

\newcolumntype{G}{>{\columncolor{LightGray}}c}

\usepackage{hyperref}

\def\beq{\begin{equation}}
\def\eeq{\end{equation}}

\def\bea{\arraycolsep .1em \begin{eqnarray}}
\def\eea{\end{eqnarray}}
\def\Tr{{\rm Tr}}

\def\eps{\epsilon}

\def\nc{N_{\rm c}}
\def\nf{N_{\rm f}}

\def\eq#1{(\ref{#1})}

\def\s0#1#2{\mbox{\small{$ \frac{#1}{#2} $}}}
\def\0#1#2{\frac{#1}{#2}}

\def\grgl{\:\hbox to -0.2pt{\lower2.5pt\hbox{$\sim$}\hss}{\raise3pt\hbox{$>$}}\:}
\def\klgl{\:\hbox to -0.2pt{\lower2.5pt\hbox{$\sim$}\hss}{\raise3pt\hbox{$<$}}\:}

\def\lsim{\mathrel{\rlap{\lower4pt\hbox{\hskip1pt$\sim$}}
    \raise1pt\hbox{$<$}}}                
\def\gsim{\mathrel{\rlap{\lower4pt\hbox{\hskip1pt$\sim$}}
    \raise1pt\hbox{$>$}}}                

\makeatletter
    \def\CT@@do@color{%
      \global\let\CT@do@color\relax
            \@tempdima\wd\z@
            \advance\@tempdima\@tempdimb
            \advance\@tempdima\@tempdimc
    \advance\@tempdimb\tabcolsep
    \advance\@tempdimc\tabcolsep
    \advance\@tempdima2\tabcolsep
            \kern-\@tempdimb
            \leaders\vrule
                    \hskip\@tempdima\@plus  1fill
            \kern-\@tempdimc
            \hskip-\wd\z@ \@plus -1fill }
    \makeatother

\begin{document}

\title{Conformal Windows  beyond Asymptotic Freedom}
\author{Andrew D.~Bond}
\email{a.bond@sussex.ac.uk}
\author{Daniel F.~Litim}
\email{d.litim@sussex.ac.uk}
\author{Gustavo Medina Vazquez}
\email{g.medina-vazquez@sussex.ac.uk}
\affiliation{\mbox{Department of Physics and Astronomy, U Sussex, Brighton, BN1 9QH, U.K.}}

\begin{abstract}
We study  four-dimensional gauge theories coupled to fermions in the fundamental and  meson-like scalars. 
All requisite beta functions are provided for general gauge group and  fermion representation.
In the regime where asymptotic freedom is absent, we determine all interacting fixed points 
using  perturbation theory up to three loop  in the gauge and two loop in the Yukawa and quartic couplings.
We  find that the conformal window of ultraviolet fixed points is narrowed-down  by finite-$N$ corrections beyond  the Veneziano limit. 
We also find a new infrared fixed point whose main features such as scaling exponents, UV-IR connecting trajectories, and phase diagram are provided. 
Both fixed points collide upon varying the number of fermion flavours $N_{\rm f}$, and conformality  is lost  through a  saddle-node bifurcation.  
We further  revisit the prospect for ultraviolet fixed points in the large $N_{\rm f}$~limit where matter field fluctuations dominate. 
Unlike at weak coupling, we do not find clear evidence for new scaling solutions even in the presence of scalar and Yukawa couplings.
 \end{abstract}

\maketitle

\tableofcontents

\section{\bf Introduction}
The seminal discovery of asymptotic freedom has established, for the first time, that local quantum field theories 
can remain well-defined and predictive up to highest energies  \cite{Gross:1973id, Politzer:1973fx}. In the language of Wilson's renormalisation group \cite{Wilson:1971bg}, asymptotic freedom corresponds to a free ultraviolet fixed point for  running couplings $(g_*=0)$. A fingerprint of asymptotic freedom is that the high-energy limit for  gauge couplings $g^2(\mu)$ is approached logarithmically slowly with increasing energy scale $\mu$,
\beq\label{AF}
g^2(\mu)\sim \frac{1}{\ln(\mu/\Lambda)}\,,
\eeq
such as in QCD. More recently, it has been noted that the high-energy limit may remain interacting, a scenario referred to as asymptotic safety.  Initially conjectured in   \cite{Bailin:1974bq},  
the availability of  interacting UV fixed points in QED- and QCD-like  theories   has raised renewed  interest recently
\cite{Litim:2014uca,
Litim:2015iea,
Bond:2016dvk,
Codello:2016muj,
Bond:2017wut,
Bond:2017sem,
Buyukbese:2017ehm,
Bond:2017tbw,
Bond:2017lnq,
Bond:2017suy,
Kowalska:2017fzw,
Abel:2017ujy,
Bond:2018oco,
Barducci:2018ysr,
Hiller:2019tvg,
Hiller:2019mou,
Bond:2019npq,
Hiller:2020fbu,
Bissmann:2020lge}.
Key ingredients for asymptotic safety  to occur
are scalar, fermionic, and vector degrees of freedom, alongside Yukawa interactions which can stabilise non-free gauge couplings \cite{Bond:2016dvk}.

Unlike in asymptotic freedom \eq{AF}, asymptotic safety in the high-energy limit 
is often characterised by a power-law-like running of the  gauge coupling $(g_*\neq 0)$, 
\beq\label{AS}
g^2(\mu)-g^2_*\sim \left(\frac{\Lambda}{\mu}\right)^A\,,
\eeq
where the scaling exponent $A$ is a universal theory-dependent number  \cite{Litim:2014uca,Bond:2017wut,Bond:2018oco}.\footnote{For exceptions to this in supersymmetry, see \cite{Bond:2017suy}.}
By now,  necessary and sufficient  conditions 
for weakly  interacting fixed points   in 4d quantum field theories are available  \cite{Bond:2016dvk,Bond:2018oco}.
Fixed points and scaling exponents have been determined 
  in simple \cite{Litim:2014uca,Bond:2017tbw,Buyukbese:2017ehm,Bond:2019npq}, semi-simple \cite{Bond:2017lnq}  and  supersymmetric   \cite{Bond:2017suy} gauge theories with matter. These ideas have also   been put forward  to UV complete the Standard Model 
\cite{Bond:2017wut,Kowalska:2017fzw,Barducci:2018ysr}, to study aspects of flavour  \cite{Hiller:2019tvg,Hiller:2019mou,Bissmann:2020lge} and to stabilise Standard Model extensions up to the Planck scale  and beyond  \cite{Hiller:2020fbu,Alkofer:2020vtb}.
Vacuum stability \cite{Litim:2015iea}, higher order interactions \cite{Buyukbese:2017ehm}, extensions away from four dimensions \cite{Codello:2016muj},  conformal windows \cite{Bond:2017lnq}, and high temperature symmetry restoration \cite{Bajc:2020gpa}  have also been addressed.
For further studies of ultraviolet fixed points, see  \cite{Martin:2000cr,Gies:2003dp,Shaposhnikov:2008xi,Gies:2013pma,Tavares:2013dga,Abel:2013mya,Intriligator:2015xxa,McDowall:2018ulq,Schuh:2018hig,Gies:2020xuh}.

In this paper, we study  interacting fixed points and their conformal windows in   QED- and QCD-like theories. 
Understanding the existence of fixed points or otherwise, particularly at low matter multiplicities,  is a crucial ingredient for  phenomenological applications in particle physics and model building \cite{Bond:2017wut,Kowalska:2017fzw,Barducci:2018ysr,Bond:2018oco,Bond:2019npq,Hiller:2019tvg,Hiller:2019mou,Bissmann:2020lge,Hiller:2020fbu,Alkofer:2020vtb}.  We also look into how conformality is lost upon increasing the number of fermion fields, and whether a new conformal regime might be found  at strong coupling. 
To these ends, we provide the general expressions for RG equations up to three loop for simple gauge theories with fermions in general irreducible representations, coupled to meson-like scalars. Specialising to $SU(\nc)$ gauge theories in the regime where asymptotic freedom is absent, 
we determine interacting fixed points 
in the Veneziano limit and beyond. We are particularly interested in the finite $N$ corrections to conformal windows, extending the work of  \cite{Bond:2017tbw} beyond the Veneziano limit. We also address the perturbativity of fixed points, and provide a comparison with QCD at electroweak energies. 

We further investigate  the appearance of a new infrared fixed point, 
which in perturbation theory becomes visible  for the first time at  three loop.
We show that it is responsible for the loss of conformality at high energies through a fixed point merger, triggered by increasing the
number of fermion species $\nf$.
Finally, following earlier conjectures \cite{PalanquesMestre:1983zy, Gracey:1996he, Holdom:2010qs} and a recent point of critique  \cite{Alanne:2019vuk}, 
we  revisit the availability of fixed points in the regime of large $\nf$ and finite $\nc$ using perturbation theory.

 The paper is organised as follows. In Sect.~\ref{Back}, we provide some background and the relevant RG equations.  We also recall the basic mechanism for fixed points in the Veneziano limit.
In Sect.~\ref{weak}, we apply our methodology to investigate ultraviolet fixed points. Results are provided for fixed points and scaling dimensions including finite $N$ corrections.
In Sect.~\ref{UV}, we analyse the UV conformal window and derive bounds  based in beta functions and on perturbation theory. We also compare findings for fixed points with perturbative QCD.
In Sect.~\ref{IR}, we point out that the lower bound of the UV conformal window arises through a fixed point merger. We provide some details about the new IR fixed point including its phase diagram, and UV-IR connecting trajectories.
In Sect.~\ref{Strong}, we revisit the many fermion limit and ask whether a strongly-coupled gauge Yukawa fixed point is supported by perturbation theory.
We present our conclusions in Sect.~\ref{Conclusions}. An appendix summarises general expressions for beta functions up to three loop.

\section{\bf Background}\label{Back}
In this section we introduce our basic models and the relevant RG beta functions.

\subsection{Family of QED- and QCD-like Theories}
We consider families of  four-dimensional Yang-Mills theory with simple gauge group $\cal G$ coupled to $\nf$ massless Dirac fermions $\psi$ and elementary mesons $H$. 
By definition, the mesons are uncharged under the gauge group and carry two flavour indices, such that they can be written as a  $\nf \times \nf$ complex matrix. The mesons interact  with fermions through a Yukawa interaction.
The theory has a global $SU(\nf)_L\times SU(\nf)_R$ flavour symmetry, and the renormalisable Lagrangian is given by
\begin{align}\label{L}
\begin{split}
L &= -\frac{1}{2} \Tr \left( F_{\mu \nu} F^{\mu \nu} \right) - \Tr \left( \bar{\psi} i \slashed D \psi \right) \\
&\mathrel{\phantom{=}} + \Tr \left( \partial_\mu H^\dagger \partial^\mu H \right)
	 + y \Tr \left( \bar{\psi}_L H \psi_R  + \psi_L H^\dagger \bar{\psi}_R \right) \\
&\mathrel{\phantom{=}}- u \Tr \left((H^\dagger H)^2 \right) - v \left(\Tr \left(H^\dagger H \right)\right)^2
\end{split}
\end{align}
where $F_{\mu \nu}$ is the field strength of the gauge bosons and $\psi = \psi_L + \psi_R$ are chiral fermions which can be separated in left-handed and right-handed components. The trace runs over the colour and flavour indices. 
The beta functions for the quantum field theory with Lagrangian \eq{L} with general compact simple gauge group $\gp$, and $\nf$ Dirac fermions in an irreducible representation $R$, are provided in App.~\ref{Appendix}. For the sake of this work, we mostly restrict ourselves to the gauge group $\gp=SU(\nc)$ with fermions in the fundamental representation, which leaves us with 
$\nc$ and $\nf$ as remaining free parameters.

Next, we introduce 't~Hooft couplings by scaling perturbative loop factors and matter field multiplicities into the definition of couplings
\begin{equation}\label{eq:alphasN}
\alpha_x=\frac{x^2 \nc}{\left(4\pi\right)^2}\,, \quad 
	\alpha_u=\frac{u \nf}{\left(4\pi\right)^2}		\,, \quad	\alpha_v=
\frac{v \nf^2}{\left(4\pi\right)^2}
\end{equation}
where $x=g,y$. Notice that the single and double trace scalar couplings scale linearly and quadratically with matter field multiplicity. Below, we choose to trade the free parameters $(\nf,\nc)$ for the set of parameters 
\beq\label{Nceps}
(\epsilon, \nc)\,,
\eeq
where $\epsilon$ is given as 
\begin{align}
 \label{eq:eps}
 \epsilon = \frac{\nf}{\nc}-\frac{11}{2}\,.
\end{align}
In the Veneziano limit, $\nc, \nf \to \infty$, the parameter \eq{eq:eps} becomes continuous and may take any value between $(-\s0{11}2,\infty)$, which would reduce the number of free parameters to  one, $\epsilon$.
The virtue of the parameter \eq{eq:eps} is that it is proportional to 
the one loop coefficient of the gauge beta function. Consequently, for
\begin{align}
\label{eq:0eps1}
0<|\epsilon|\ll1
\end{align} 
strict perturbative control  is achieved. In practice, at finite $N$, \eq{eq:eps} can no longer be taken as continuous. Still, we continue to assume that $\epsilon$ can be taken sufficiently small to achieve perturbative control. 

For $\eps<0$, the theories \eq{L} are asymptotically free and  we refer to them as ``QCD-like''. In the infrared, they either display confinement and chiral symmetry breaking, or, provided that  $\eps$ is small enough, a regime with IR conformality due to a Banks-Zaks fixed point. 

Conversely, for $\eps>0$, asymptotic freedom is absent and the theories \eq{L} can be viewed as non-abelian versions of QED. It has been shown previously that these theories can then develop weakly coupled asymptotically safe UV  fixed points. The models remain well-defined and predictive up to highest energies, and offer several scenarios in the low-energy regime.  In the limit of small couplings the theories can become ``QED-like'' in that gluons and fermions can become infrared free, very much like in  massless QED.  In turn, couplings may also grow towards the IR, in which case the models are once more "QCD-like", with  either  confinement and chiral symmetry breaking,  or conformality in  the deep IR. 

In this work, we are mostly interested in  regimes where $\eps>0$.

\subsection{Renormalisation Group}

The renormalisation group beta functions for this class of theories are formally known in the $\mathrm{\overline{MS}}$ scheme \cite{Machacek:1983tz,Machacek:1983fi,Machacek:1984zw,Pickering:2001lfn,Luo:2002ti}, and can be extracted either manually or with the help of  suitable codes \cite{Litim:2020jvl}. In perturbation theory,  we  write them as
\begin{align}
\beta_i \equiv \frac{d\alpha_i}{d\ln \mu} = \beta_i^{(1)} + \beta_i^{(2)} + \beta_i^{(3)} + \dots\,,
\end{align}
where $\beta_i^{(n)}$ denotes the $n$-th loop contribution, and $i$ any of $\{g,y,u,v\}$.
Below, we investigate approximations which retain different loop orders of couplings in different 
beta functions. 
Following  \cite{Litim:2014uca,Bond:2017tbw}, we   introduce the notation ``{\texttt{klm}}"
to denote a perturbative approximation of beta functions 
which retains \texttt{k} ~loop orders in the gauge beta 
function, \texttt{l}~loops in the Yukawa, and \texttt{m}~loops in the scalar beta functions.

Results for  beta functions of the theory \eq{L} 
for general gauge group and fermion representation
are summarised in  App.~\ref{Appendix}.
Here, we state them for $SU(\nc)$ gauge theories with $\nf $ fermions in the fundamental representation up to the 
  {\texttt{322}} approximation. The next complete  order of approximation, which would be \texttt{433}, is 
  presently unavailable though 
  some partial results already exist.\footnote{In the scalar-Yukawa sector $(\alpha_g=0)$,  
 general results for beta functions up to  three loops have been made available in \cite{Steudtner:2021fzs}. 
Further, novel computational techniques have recently extended general 
expressions  in the scalar sector $(\alpha_g=0=\alpha_y)$   up to four  \cite{Steudtner:2020tzo} 
and six loop \cite{Bednyakov:2020ifb}, and in the gauge sector up to four loop \cite{Bednyakov:2021qxa} 
 ({\texttt{432}} approximation).}
  
In the   {\texttt{322}} approximation considered here,
   the first three terms of the gauge beta function and in terms of \eq{eq:alphasN} take the following form
\begin{align}
    \label{eq:beta_g}
\begin{split}
    \textstyle \beta_g^{(1)} &= \textstyle \frac{4\epsilon}{3}\alpha_g^2 
    \\
    \textstyle \beta_g^{(2)} &= \textstyle 
		\left(25+\frac{26 \epsilon }{3}-\frac{11 + 2 \epsilon}{\nc^2}\right)\alpha_g^3 
		- 2 \left(\frac{11}{2}+\epsilon\right)^2 \alpha_g^2 \alpha_y \\
        \textstyle \beta_g^{(3)} &= \textstyle 
		\Bigl(\frac{6309+954 \epsilon -224 \epsilon ^2}{54}
		+\frac{11 (11+2\epsilon)\left(\epsilon-3\right)}{18 \nc^2} \Bigr)\alpha_g^4
		 \\
	&\mathrel{\phantom{=}} \textstyle
		-\frac{11 + 2 \epsilon}{4 \nc^4}\alpha_g^4 
		-  \frac{3}{8} \left(9 - \frac{1}{\nc^2}\right) (11 + 2 \epsilon)^2 \alpha_g^3 \alpha_y 
		\\
	&\mathrel{\phantom{=}} \textstyle
		+ \frac{1}{4}(11 + 2 \epsilon)^2 (3 \epsilon +20)\alpha_g^2 \alpha_y^2
     \end{split}
\end{align}
For the Yukawa coupling, we find
\begin{align}
    \label{eq:beta_y}
\begin{split}
    \textstyle \beta_y^{(1)} &= \textstyle (13+2 \epsilon) \alpha_y^2-6\left(1-\frac{1}{\nc^2}\right) \alpha_g \alpha_y\\
        \textstyle \beta_y^{(2)} &= \textstyle 
		-\frac{1}{8} \left( (11 + 2 \epsilon)(2\epsilon + 35) -\frac{32}{\nc^2} \right)\alpha_y^3 
		\\
	&\mathrel{\phantom{=}} \textstyle
		+ \left(1-\frac{1}{\nc^2}\right) (8 \epsilon +49) \alpha_g \alpha_y^2 
				\\
        &\mathrel{\phantom{=}} \textstyle 
		+\frac{1}{6}\left(1 -\frac{1}{\nc^2}\right)\left((20\epsilon -93) +\frac{9}{\nc^2}\right) \alpha_g^2 \alpha_y 
		\\
        &\mathrel{\phantom{=}} \textstyle 
		-4\left((11 + 2 \epsilon) +\frac{4}{(11 + 2 \epsilon)\nc^2}\right)  \alpha_u \alpha_y^2 
		\\
        &\mathrel{\phantom{=}} 
        \textstyle 
		+ 4\left(1+\frac{4}{\left(11 + 2 \epsilon\right)^2\nc^2}\right)\alpha_u^2 \alpha_y 
		\\
        &\mathrel{\phantom{=}} \textstyle 
        +\frac{64}{\left(11 + 2 \epsilon\right)^2\nc^2 }\alpha_u \alpha_v \alpha_y
        -\frac{16}{(11 + 2 \epsilon)\nc^2}\alpha_v \alpha_y^2
		\\
    &\mathrel{\phantom{=}} \textstyle 
		+ \frac{16}{\left(11 + 2 \epsilon\right)^2\nc^2} \left(1+\frac{4}{\left(11 + 2 \epsilon\right)^2\nc^2 }\right) \alpha_v^2 \alpha_y
 \end{split}
\end{align}
For the scalar single-trace interaction we obtain
\begin{align}
    \label{eq:beta_u}
\begin{split}
   \textstyle \beta_u^{(1)} &= \textstyle 8 \alpha_u^2+4 \alpha_u \alpha_y- (11 + 2 \epsilon)\alpha_y^2 
		+\frac{96\alpha_u \alpha_v}{\left(11 + 2 \epsilon\right)^2\nc^2 }
 \\
        \textstyle \beta_u^{(2)} &= \textstyle 
		-24 \left(1 +\frac{20}{\left(11 + 2 \epsilon\right)^2 \nc^2}\right) \alpha_u^3 
		- 16 \alpha_y  \alpha_u^2 \\
	&\mathrel{\phantom{=}} \textstyle
		- \frac{1408\alpha_u^2 \alpha_v}{\left(11 + 2 \epsilon\right)^2 \nc^2} 
		-3 \left(11 + 2 \epsilon\right) \alpha_y^2 \alpha_u \\
	&\mathrel{\phantom{=}} \textstyle
		- \frac{32 \alpha_u \alpha_v^2}{\left(11 + 2 \epsilon\right)^2 \nc^2}\left( 5 + \frac{164}{\left(11+2\epsilon\right)^2 \nc^2} \right) \\
        &\mathrel{\phantom{=}} \textstyle
		+10 \left(1 -\frac{1}{\nc^2}\right) \alpha_g \alpha_y \alpha_u 
		- \frac{192\alpha_y \alpha_u \alpha_v}{\left(11 + 2 \epsilon \right)^2 \nc^2}   \\
	&\mathrel{\phantom{=}} \textstyle
		-2 \left(11 + 2 \epsilon\right) \left(1 - \frac{1}{\nc^2}\right) \alpha_g \alpha_y^2 \\
        &\mathrel{\phantom{=}} \textstyle 
		+ \frac{16}{\left(11 + 2 \epsilon\right) \nc^2} \alpha_y^2 \alpha_v
		+ \left(11 + 2 \epsilon \right) \alpha_y^3\,.
  \end{split}
\end{align}
Finally, for the double-trace self-interaction, we have
\begin{align}
    \label{eq:beta_v}
\begin{split}
    \textstyle \beta_v^{(1)} &= 
		\textstyle  12 \alpha_u^2+16 \alpha_u \alpha_v+4 \alpha_v \alpha_y\\
        &\mathrel{\phantom{=}} \textstyle 
		+4\left(1+\frac{16}{\left(11 + 2 \epsilon\right)^2\nc^2}\right)\alpha_v^2 
\\
        \textstyle \beta_v^{(2)} &= \textstyle 
		-\frac{96\alpha_v^3}{\left(11 + 2 \epsilon\right)^2 \nc^2} \left( 3 + \frac{28}{\left(11 + 2 \epsilon\right)^2 \nc^2} \right)  
		\\
       &\mathrel{\phantom{=}} \textstyle 
		-8 \left( 1 + \frac{16}{\left(11 + 2 \epsilon\right)^2 \nc^2} \right) \alpha_y \alpha_v^2 
		-\frac{1408\alpha_u \alpha_v^2}{\left(11 + 2 \epsilon\right)^2 \nc^2} \\
        &\mathrel{\phantom{=}} \textstyle
		- \left(11 + 2 \epsilon \right) (3\alpha_v -4\alpha_u)\alpha_y^2 \\
        &\mathrel{\phantom{=}} \textstyle 
		-8 \left(5 + \frac{164}{\left(11 + 2 \epsilon\right)^2 \nc^2} \right) \alpha_u^2 \alpha_v \\
        &\mathrel{\phantom{=}} \textstyle 
		+10 \left(1-\frac{1}{\nc^2}\right)\alpha_g \alpha_y \alpha_v 
		- 32 \alpha_y \alpha_u \alpha_v 
		\\
        &\mathrel{\phantom{=}} \textstyle 
		- 24 \alpha_y \alpha_u^2 
		+ \left(11 + 2 \epsilon\right)^2 \alpha_y^3 -96 \alpha_u^3 \,.
 \end{split}
\end{align}
A few technical comments are in order. 
The \texttt{210} approximation determines the coordinates of the Banks-Zaks and the gauge-Yukawa fixed point reliably for small enough $\eps$. This continues to be true in the presence of finite $N$ corrections, the main reason being that the scalar couplings do not contribute to the gauge-Yukawa subsector at this loop order. 
The scalar couplings only start contributing to the running of the gauge coupling at the fourth loop order.  At the \texttt{322} approximation, the main addition beyond the Veneziano limit is an enhanced entanglement between the Yukawa and the quartic sectors. Specifically, we find new finite $N$ contributions  to $\beta_y$ 
proportional to $\alpha_y\alpha_u\alpha_v$, $\alpha_y^2\alpha_v$, and $\alpha_y\alpha_v^2$. Similarly, the new contributions  to $\beta_u$ are proportional to 
$\alpha_u\alpha_v$, $\alpha_u\alpha^2_v$, $\alpha_u^2\alpha_v$, $\alpha_y^2\alpha_v$ and $\alpha_y\alpha_u\alpha_v$, while  
$\beta_v$ receives additional contributions proportional to $\alpha_v^3$ and $\alpha_u\alpha^2_v$. Below, we quantify the effect of the additional terms on fixed points and the size of conformal windows.

\section{\bf  Weakly Coupled Fixed Points}\label{weak}
In this section, we discuss the systematics of the weak coupling expansion and the results for
fixed points and  scaling exponents.

\subsection{Fixed Points}\label{sub:systematics_of_weakly_coupled_fixed_points}
The theory described above admits a non-trivial weakly coupled fixed point in the Veneziano limit. The size of couplings is controlled by the perturbative parameter $\epsilon$ \eqref{eq:eps}, and the fixed point can be expressed as a series expansion:
\begin{align}
\label{eq:fpx}
\alpha_x^* &= c_{x,1}\epsilon + c_{x,2}\epsilon^2 +O\left(\epsilon^3\right)
\end{align}
where $x=\{g,y,u,v\}$. The coefficients of these series can be found by systematically solving for the stationary point of the beta functions order by order. To consistently determine these coefficients, the loop-order expansion is superseded by an expansion in powers of $\epsilon$, giving rise to the power counting scheme put forward in ref. \cite{Litim:2014uca}. It will now be shown that this fixed point can persist qualitatively unchanged once finite $\nc$ corrections are accounted for.

Away from the Veneziano limit, $\nc$ is finite and must be specified, therefore the theory is described by two free parameters, $\epsilon$ and $\nc$. Although new terms that depend on $\nc$ appear in the beta functions, they just amount to a shift in the coefficients of those equations. To establish that a weakly coupled fixed point is still viable, consider the replacement $\alpha_x \to f_x(\nc)\epsilon$, where it is required that the couplings are proportional to $\epsilon$, which is finite but small enough such that perturbation theory is still applicable, and the unknown function $f_x$ captures the dependence on $\nc$. Inserting this into the beta functions reveals that the corrections appearing in the finite $\nc$ regime are of the same or subleading order in $\epsilon$ as those already present in the Veneziano limit. In particular, no term of constant order $\epsilon^0$ appears in the beta functions. Thus, the expansion in small $\epsilon$ still holds, with finite $\nc$ corrections modifying the coefficients of the expression, which may now be written as:
\begin{align}
\label{eq:fp_gen}
\alpha_x^* &= c_{x,1}\, f_{x,1}(\nc)\,\epsilon + c_{x,2}\, f_{x,2}(\nc)\,\epsilon^2 +O\left(\epsilon^3\right)
\end{align}
where the numbers $c_{x,i}$ from the Veneziano limit have been factored out from the coefficients of the series. The general functions $f_{x,i}(\nc)$ can be computed in the same manner as the $c_{x,i}$ were previously obtained. Although one could worry that higher loop-order terms could modify the leading order functions $f_{x,i}$, note that such terms are accompanied by increasing powers of the couplings, resulting in terms that are subleading in $\epsilon$. Therefore, the coefficients of the series expansion, both $c_{x,i}$ and $f_{x,1}$ are not modified by higher order terms by virtue of the perturbative expansion. Moreover, as with any perturbative expansion, the higher order coefficients should not increase too rapidly as to spoil the convergence of the series. This remains to be checked explicitly once the functions $f_{x,i}$ are computed. Finally, note that if the limit $\lim_{\nc\to\infty} f_{x,i}(\nc)=1$ can be taken continuously, then the fixed points found in both cases are continuously connected.

The functions $f_{g,1}$ and $f_{y,1}$ encapsulating the finite $\nc$ corrections can be computed within the \texttt{211} approximation in a straightforward manner, yielding
\begin{align}\label{eq:f_g1}
\begin{split}
f_{g,1}(\nc) = \frac{\nc^2}{\nc^2 -\frac{110}{19}} \,, \quad
f_{y,1}(\nc) = \frac{\nc^2 -1}{\nc^2 -\frac{110}{19}} \,.
\end{split}
\end{align}
These simple functions provide three important insights on how finite $\nc$ corrections are modifying the fixed point. First, note that the limit $\lim_{\nc\to\infty}f_{x,1}(\nc)=1$ is well defined, such that the fixed point of the Veneziano limit can be continuously extended into the finite $\nc$ regime. Second, these functions are monotonically increasing as $\nc$ decreases, indicating that a smaller number of gauge fields leads to a less perturbative fixed point, with the functions reaching values close to 2.8 and 2.5 at $\nc=3$, respectively. Third, the denominator effectively sets a lower boundary on the conformal window of the theory, implying that a fixed point is available for theories with $\nc\geq3$. Below this number, the expressions diverge and the perturbative $\epsilon$ expansion would break-down. Finally, we highlight that no assumption has been made about the value of $\nc$, with the functions \eqref{eq:f_g1} 
being exact expressions valid for general $\nc$.

Closed analytic expressions can similarly be found for all the other functions up to order \texttt{322}, such that the coefficients of the fixed point expressions \eqref{eq:fp_gen} are completely determined to order $\epsilon^2$. They all show a qualitatively similar behaviour, being positive, finite and monotonically increasing for decreasing $\nc$ up to $\nc\geq3$. Thus, the finite $\nc$ corrections only amount to a shift in the coefficients of the series expansion of the fixed point, making the fixed point slightly more strongly coupled for small $\nc$. Although this could, in principle, push the solution out of the perturbative domain, it will be shown in further sections that perturbative solutions are still available even at small values of $\nc$ and finite $\epsilon$.

Although these functions $f_{x,i}$ show a simple behaviour, their full form is rather complicated, arising as roots of fourth-order polynomials. Their full expressions partly obscure the relevant physics. Thus, for the sake of readability, approximated expressions are provided in here of the same form of \eqref{eq:f_g1}, which accurately capture the behaviour of the original functions, such that the fixed point coordinates can be written as
\begin{widetext}
\begin{align}\label{eq:ag_nc}
\begin{split}
   \alpha_g^* &= 
    \frac{26}{57}\frac{\nc^2}{\nc^2 -\frac{110}{19}}\epsilon 
   + \frac{23(75245-13068\sqrt{23})}{370386}\frac{\nc^2 +6.632}{\nc^2 -7.659}\epsilon^2 
\\
    \alpha_y^* &=
    \frac{4}{19}\frac{\nc^2 -1}{\nc^2 -\frac{110}{19}}\epsilon 
    + \frac{43549-6900\sqrt{23}}{20577} \frac{\nc^2 +5.509}{\nc^2 -7.508}\epsilon^2 
\\
   \alpha_u^* &= 
    \frac{\sqrt{23} -1}{19}\frac{\nc^2 -0.9737}{\nc^2 -\frac{110}{19}}\epsilon 
    + \frac{365825\sqrt{23} -1476577}{631028}\frac{\nc^2 +5.534}{\nc^2 -7.535}\epsilon^2 
\\
    \alpha_v^* &= 
    -\frac{2\sqrt{23} -\sqrt{20+6\sqrt{23}}}{19}\frac{\nc^2 -0.9474}{\nc^2 -\frac{110}{19}}\,\epsilon 
    - \left(\frac{321665}{13718\sqrt{23}} -\frac{27248}{6859} +\frac{\frac{33533}{6859} -\frac{452563}{13718\sqrt{23}}}{\sqrt{20+6\sqrt{23}}}\right) 
    \frac{\nc^2 +4.214}{\nc^2 -7.347}\, \epsilon^2 
    \end{split}
\end{align}
\end{widetext}
up to terms of order $O\left(\epsilon^3\right)$ and with associated error terms at each order. Neglecting $1/\nc$ corrections, expressions reduce to those given earlier  in \cite{Bond:2017tbw}. In these expressions, numbers written in natural form are exact, while those in decimal form were determined from fitting the model to the full functions. As previously stated, the finite $\nc$ corrections do not induce any sign changes and merely increase the magnitude of the coefficients as $\nc$ decreases. Thus, the fixed point from the Veneziano limit can be extended into the finite $\nc$ regime at up to next-to-next-to-leading order while remaining qualitatively unchanged.

Before continuing, a short word about the approximation is due now. The model was fitted using the values of the exact functions in the range $\nc\in [3,100]$. The largest of the Mean Squared Errors of these fitted models is $MSE=4.53\times10^{-5}$, amounting to a cumulative Sum of Squared Errors of $SSE=4.44\times10^{-3}$. This indicates that the model successfully captures most of the variation of the original functions $f_{x,i}$ with small associated errors. 

Notice that the coefficients in the numerator of $\alpha_u^*$ and $\alpha_v^*$ in \eqref{eq:ag_nc}  at order $\epsilon$ are both   close to unity, while it is exactly unity for $\alpha_y^*$. It has been verified that artificially fixing them to unity results in a poorer approximation, with larger errors in the small $\nc$ region, where most of the variation occurs. Since the task is to approximate an exact function in the absence of random noise, there is no bias-variance trade-off and the model with the least errors is preferred. The same goes for the denominators of the subleading $f_{x,2}$ functions. Moreover, we have made sure that a fair compromise between the magnitude of errors at low $\nc$ and high $\nc$ is achieved, such that the approximated functions are equally valid from $\nc=3$ up to the Veneziano limit.

\subsection{Vacuum Stability}
\label{sec:scalar}
Further to couplings remaining finite in the UV, the scalar potential of the renormalised theory should be bounded from below to prevent the vacuum state from being unstable. The conditions for vacuum stability in the present theory take the form \cite{Litim:2015iea}
\begin{align}
\begin{split}
\alpha_u^* &> 0	\quad	\alpha_u^* + \alpha_v^*  \geq 0\\
\alpha_u^* &<0	\quad	\alpha_u^* + \frac{\alpha_v^*}{\nf}  \geq 0
\end{split}
\end{align}
Since the scalar coupling $\alpha_u$ is positive at the identified fixed point, it is the first condition that needs to be checked. A previous study in the Veneziano limit showed that passing from the {\texttt{321}}  to the {\texttt{322}}  approximation, the subleading term of the potential appears with the opposite sign, triggering an instability of the vacuum above a maximum value $\epsilon_{\text{max}}$. Notice also that this happens close to where a fixed point merger limits the conformal window \cite{Bond:2017tbw}. 

For this reason, it is instructive to compare both approximations under the influence of finite $\nc$ corrections and verify whether limitations persist. We introduce the notation for the scalar potential at approximation {\texttt{klm}}  as
\begin{align}
U^*\big|_{\texttt{klm}}  &=    ( \alpha_u^* + \alpha_v^*)\big|_{\texttt{klm}} \,.
\end{align}
Vacuum stability $U^*>0$ is then checked at approximation order {\texttt{321}}  and {\texttt{322}}. We find
\begin{align}
\begin{split}
    U^*\big|_{\texttt{321}}  &= 0.0625 \left(\frac{\nc^2 -1.032}{\nc^2 -5.790}\right)\epsilon \\
&\mathrel{\phantom{=}}
		+0.1535 \frac{\left(\nc^2 -7.491\right)\left(\nc^2 +5.307\right)}{\left(\nc^2 -7.497 \right)\left(\nc^2 -7.496\right)} \, \epsilon^2 
\end{split}\\
\label{eq:vac322}
\begin{split}
    U^*\big|_{\texttt{322}} &= 0.0625 \left(\frac{\nc^2 -1.032}{\nc^2 -5.790}\right)\epsilon \\
&\mathrel{\phantom{=}}
		-0.1915 \frac{\left(\nc^2 -8.169\right)\left(\nc^2 +1.380\right)}{\left(\nc^2 -7.535 \right)\left(\nc^2 -7.347\right)} \, \epsilon^2
\end{split}
\end{align}
Factoring out the numeric values of the coefficients in the Veneziano limit makes it evident that the finite $\nc$ corrections do not induce a sign change in the subleading term for any value $\nc\geq3$. Therefore, vacuum instability continues posing a constraint in the maximum value of $\epsilon$ at approximation order \texttt{322}. Moreover, the coefficient of the subleading term grows faster than the one of the leading term for decreasing values of $\nc$. This defines a restriction on the conformal window as a function of $\nc$, $\epsilon_{\text{max}}=\epsilon_{\text{max}}(\nc)$, and indicates that reducing the number of gauge bosons can narrow the conformal window of the theory.

\subsection{Scaling Exponents}
The universal scaling exponents as in \eq{AS} characterise the RG flow close to the fixed point. They can be identified from the eigenvalues of the stability matrix $M_{ij} = \partial \beta_i/\partial \alpha_j\bigr|_*$ evaluated at the fixed point. Taking into account the finite $\nc$ terms, approximate expressions for the scaling exponents are
\begin{widetext}
\begin{align}
\begin{split}
\label{eq:eig1}
    \vartheta_1\big|_{\texttt{322}} &= -\frac{104}{171}\frac{\nc^2 }{\nc^2 -\frac{110}{19} }\epsilon^2 
+\frac{2296}{3249}\frac{\nc^2\left(\nc^2 -1.136\right)}{\left(\nc^2 -5.789\right)^2}\epsilon^3 
\\
    \vartheta_2\big|_{\texttt{322}} &= 
		\frac{52}{19}\frac{\nc^2 -1}{\nc^2 -\frac{110}{19}}\epsilon 
        + \frac{136601719-22783308\sqrt{23}}{4094823}\left(\frac{\nc^2 -0.458}{\nc^2 -6.219}\right)^2\epsilon^2 
\\
   \vartheta_3\big|_{\texttt{322}} &= \frac{8\sqrt{20+6\sqrt{23}}}{19}\frac{\nc^2 -1.148}{\nc^2 -\frac{110}{19}}\epsilon 
		+ \frac{2\sqrt{2}(50059110978+10720198219\sqrt{23})}{157757\left(10+3\sqrt{23}\right)^{9/2}} 
        \frac{\left(\nc^2 +0.5327\right)\left(\nc^2 +18.84\right)}{\left(\nc^2 -6.294\right)^2}\epsilon^2 
\\
   \vartheta_4\big|_{\texttt{322}} &= \frac{16\sqrt{23}}{19}\frac{\nc^2 -0.9250}{\nc^2 -\frac{110}{19}}\epsilon 
 +\frac{4(68248487\sqrt{23}-255832864)}{31393643}
\left(\frac{\nc^2 -0.6640}{\nc^2 -6.301}\right)^2\epsilon^2 \,.
\end{split}
\end{align}
\end{widetext}
Neglecting $1/\nc$ corrections, expressions reduce to  the results   in \cite{Bond:2017tbw}.  The effect of finite $\nc$ corrections are qualitatively similar to what we observed for the  fixed point couplings. Here, the subleading expressions are fitted using a model with higher powers of $\nc$ to best fit the exact result. The approximation with the largest error has $MSE=7.44\times10^{-5}$ and $SSE=7.00\times10^{-3}$, indicating once again a good fit. The  finite $\nc$ corrections do not induce a sign change and are positive and finite for $\nc\geq3$. 

Notice that the subleading corrections to the relevant scaling exponent $\vartheta_1$ appear with the opposite sign  \cite{Bond:2017tbw}. This hints at the possibility for a strict cancellation, which does arise at a fixed point merger. If so, it would occur at lower $\epsilon$ for lower  $\nc$. We will expand more on this   in Sect.~\ref{IR} 

\begin{table}[b]
\aboverulesep = 0mm
\belowrulesep = 0mm
	\addtolength{\tabcolsep}{1pt}
	\setlength{\extrarowheight}{1pt}
\begin{tabular}{`cccc`}
 \toprule
  \rowcolor{Yellow}
{\bf $\nc$} &\bf $\epsilon_{\rm max}$ & \bf $N_{\rm Fmin}$ & \bf $N_{\rm Fmax}$\\
\midrule
 $\infty$ &0.326\, \,  & $\s0{11}2\nc$ & $(\s0{11}2+\eps_{\rm max})\nc$ \\
\rowcolor{LightGray}
 $7$ &0.3044 & 39 & 40 \\
 $5$ &0.2830 & 28 & 28 \\
\rowcolor{LightGray}
$3$ & 0.2278 & 17 & 17\\
\bottomrule
\end{tabular}
\caption{Bounds and selected integer value solutions $(\nc,\nf)$ within the UV conformal window of Fig.~\ref{fig:cw1}.}
\label{tbl:valuesNC}
\end{table}

\begin{figure}[t]
   \includegraphics[width=.8\linewidth]{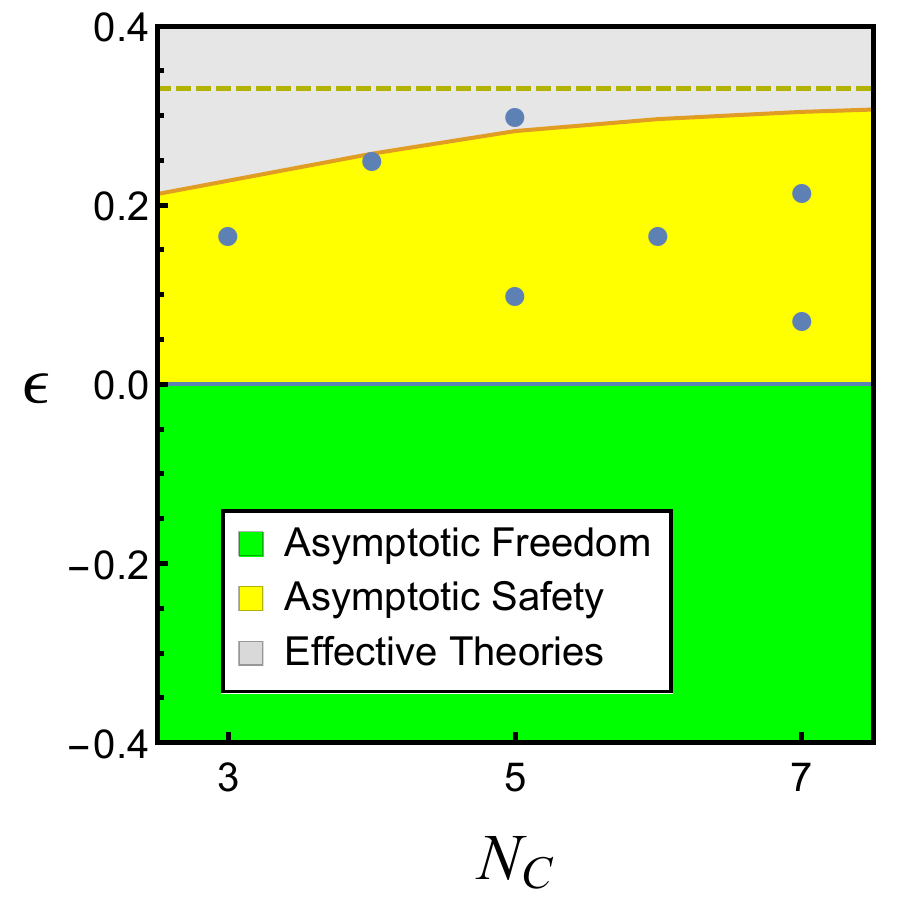}
    \caption{Conformal window from fixed points at NNLO.
    The upper boundary is given by vacuum stability. Markers represent values of $\epsilon$ for integer values of $\nc$. The boundary in the Veneziano limit is indicated by the dashed line.}
    \label{fig:cw1}
\end{figure}

\section{\bf  Interacting UV Fixed Points}\label{UV}
In this section, we evaluate the finite $N$ corrections to the UV conformal window.
We also investigate the perturbativity of the fixed point, in particular at the upper boundary.

 \label{sec:cw}

\subsection{UV Conformal Window}
From the fixed point analysis we can start to identify constraints in the parameter space of the theory. The first one comes from \eqref{eq:f_g1}, which grows large as $\nc$ decreases and effectively sets a lower boundary for the number of fields $N_{c, \rm min}=3$. The second arises through vacuum instability \eqref{eq:vac322} with the constraint in $\epsilon$ given as a function of $\nc$; in the Veneziano limit we reproduce the result found in \cite{Bond:2017tbw}. We can identify a third one where the relevant eigenvalue becomes irrelevant due to a sign change at next-to-leading order \eqref{eq:eig1}, however this constraint is not as strong as the one arising from vacuum stability.
The key result of this section is that as $\nc$ decreases, $\epsilon_{\rm max}(\nc)$ decreases as well; in other words, the conformal window becomes more narrow as the number of fields $\nc$ decreases. This is illustrated in Figure \ref{fig:cw1} where we have plotted the conformal window of the theory, bounded by vacuum stability, reflecting the full $\nc$ dependence. For comparison the boundary obtained in the Veneziano limit is also plotted (dashed line). We also list the values of $\epsilon_{\rm max}$ for a few interesting cases of finite $\nc$ in Table \ref{tbl:valuesNC}. We notice that as $\nc$ decreases, the minimum and maximum number of fermions required is also constrained, and vice versa. Note also that $\epsilon_{\rm max}$ remains relatively small at finite $\nc$, meaning that the perturbative expansion is still justified even if we saturate the bound on the number of fermions allowed at each $\nc$.

\begin{table*}[t]
\aboverulesep = 0mm
\belowrulesep = 0mm
	\addtolength{\tabcolsep}{1pt}
	\setlength{\extrarowheight}{1pt}
\begin{tabular}{`cc cc cc cc cc`}
 \toprule
 \rowcolor{Yellow}
 {}\ \ \ \bf Couplings\ \ \ 
 &\multicolumn{9}{c`}{\ \ \bf Orders in Perturbation Theory\ \ }
 \\
\midrule
\rowcolor{LightGray}     
 $\beta_{\rm gauge}$ &2&2&2&2&2&3&3&3& 3\\
\rowcolor{white}
 $\beta_{\rm  Yukawas}$&1&1&1&2&2&1&1&2&2\\
\rowcolor{LightGray}
 $\beta_{\rm   quartics}$ &0&1&2&1&2&1&2&1&2\\
\midrule
$\bm{\epsilon_{\rm subl.}|_\text{\footnotesize Veneziano}}$ 
&$\ 1.048^a\ $
&$\ 1.048^a\ $
&$\ 0.116^c\ $
&$\ 3.112^b\ $
&$\ 0.208^c\ $
&$\ 0.027^b\ $
&$\ 0.027^b\ $
&$\ 0.117^b\ $
&$\ 0.087^c\ $\\
\rowcolor{LightGray}
$\bm{\epsilon_{\rm subl.}|_{\nc=7}}$ 
&$\ 0.898^a\ $
&$\ 0.898^a\ $
&$\ 0.104^c\ $
&$\ 3.183^b\ $
&$\ 0.198^c\ $
&$\ 0.021^b\ $
&$\ 0.021^b\ $
&$\ 0.105^b\ $
&$\ 0.077^c\ $\\
$\bm{\epsilon_{\rm subl.}|_{\nc=5}}$ 
&$0.762^a$
&$0.762^a$
&$0.092^c$
&$3.256^b$
&$0.189^c$
&$0.017^b$
&$0.017^b$
&$0.094^b$
&$0.068^c$\\
\bottomrule
\end{tabular}
\caption{Maximal values
$\eps_{\rm subl.}$ for the Veneziano parameter $\eps$ up until which asymptotic safety is realised in the Veneziano limit and in the finite $\nc$ regime. Limits arise due to $a)$ strong coupling, $b)$ fixed point mergers, or $c)$ vacuum instability. Data for $\epsilon_{\rm subl.}$ in the Veneziano limit taken from \cite{Bond:2017tbw}.}
\label{tbl:eps}
\end{table*}

\subsection{Bounds from Perturbation Theory}\label{BFPT}

We can probe the impact of higher loop orders by considering subleading effects in the beta functions, following  \cite{Bond:2017tbw}, while keeping the  $\nc$ dependence explicit.
First we would like to get some idea about how higher loops orders of the beta functions would be behaving at the fixed point we are studying. 
To that end, we substitute \eqref{eq:ag_nc}  to order $O\left(\epsilon\right)$ into the beta functions, taking a series expansion in $\epsilon$ and evaluating at the fixed point. The leading non-vanishing terms are:
\begin{align}
\label{eq:betas_shift_211}
\begin{split}
    \beta_g^{(3)}|_{\texttt{211}} &= 2.477 \left(\frac{\nc^2 +24.51}{\nc^2 -8.390}\right) \epsilon^4\\
    \beta_y^{(2)}|_{\texttt{211}} &= -0.4934 \left(\frac{\nc^2 +11.11}{\nc^2 -7.871}\right) \epsilon^3\\
    \beta_u^{(2)}|_{\texttt{211}} &= 0.2581 \left(\frac{\nc^2 +9.359}{\nc^2 -7.813}\right) \epsilon^3\\
    \beta_v^{(2)}|_{\texttt{211}} &= 0.9925 \left(\frac{\nc^2 +9.341}{\nc^2 -7.814}\right) \epsilon^3
\end{split}
\end{align}
Negative shifts to the beta functions will, in general, lead to a wider conformal window at a higher loop order \cite{Bond:2017tbw}. This can be explained by recalling that the interacting fixed point is generated by a positive one loop term and a negative two loop term. Any higher order terms with a negative sign will drive the zero to smaller coupling values. On the contrary, terms with a positive sign can shift the zero towards larger values. Large enough positive values could even destabilize the fixed point by preventing the cancellation from occuring in the first place. Overall, this gives us a qualitative picture of what the tendency is at higher loop orders. Likewise, we can do the same for the result at order $\epsilon^2$ and obtain:
\begin{align}
\label{eq:betas_shift_322}
\begin{split}
    \beta_g|_{\texttt{322}} &= 10.24 \left(\frac{\nc^2 +47.13}{\nc^2-8.732}\right) \epsilon^5\\
    \beta_y|_{\texttt{322}} &= -1.713 \left(\frac{\nc^2 +28.60}{\nc^2-8.527}\right) \epsilon^4\\
    \beta_u|_{\texttt{322}} &= 1.696 \left(\frac{\nc^2 +22.47 }{\nc^2-8.435}\right) \epsilon^4\\
    \beta_v|_{\texttt{322}} &= 7.237 \left(\frac{\nc^2 +21.23}{\nc^2-8.420}\right) \epsilon^4
\end{split}
\end{align}
The result is qualitatively the same, with a majority of positive signs suggesting an overall destabilizing effect for the UV fixed point. This, in turn, means that the conformal window is likely to be further constrained. We note that the $\nc$ functions do not change this behaviour and merely just scale the magnitude of the coefficients. We now go a step further and try to compute quantitative bounds. 

The influence of higher loop order terms on the conformal window can be quantitatively estimated using partial information of these terms \cite{Bond:2017tbw}. First, note that when couplings take fixed point values, $\alpha_i \propto \epsilon$, then $\epsilon$ effectively becomes the small parameter in the beta functions, meaning that we can order the beta function by powers of $\epsilon$ (this is indeed how the fixed point is computed in the first place). At the $n$-th loop order, the beta functions are of order $\beta \sim O\left(\epsilon^{n+1}\right)$. Terms proportional to $\epsilon^{m}$ with $m>n+1$ are subleading in the weak coupling regime and were neglected when computing the fixed point expressions. Note that at an interactive fixed point, the subleading terms would cancel with higher order terms that would be included at the next loop order. We now put forward an approach in which we keep all these subleading terms, treating beta functions as if they were exact at the given approximation order. Constraints on the conformal window can then be computed from the fixed points of this system. We refer to constraints obtained in this approach as $\epsilon_{\rm subl}$.

\begin{figure}[b]
\includegraphics[width=.8\linewidth]{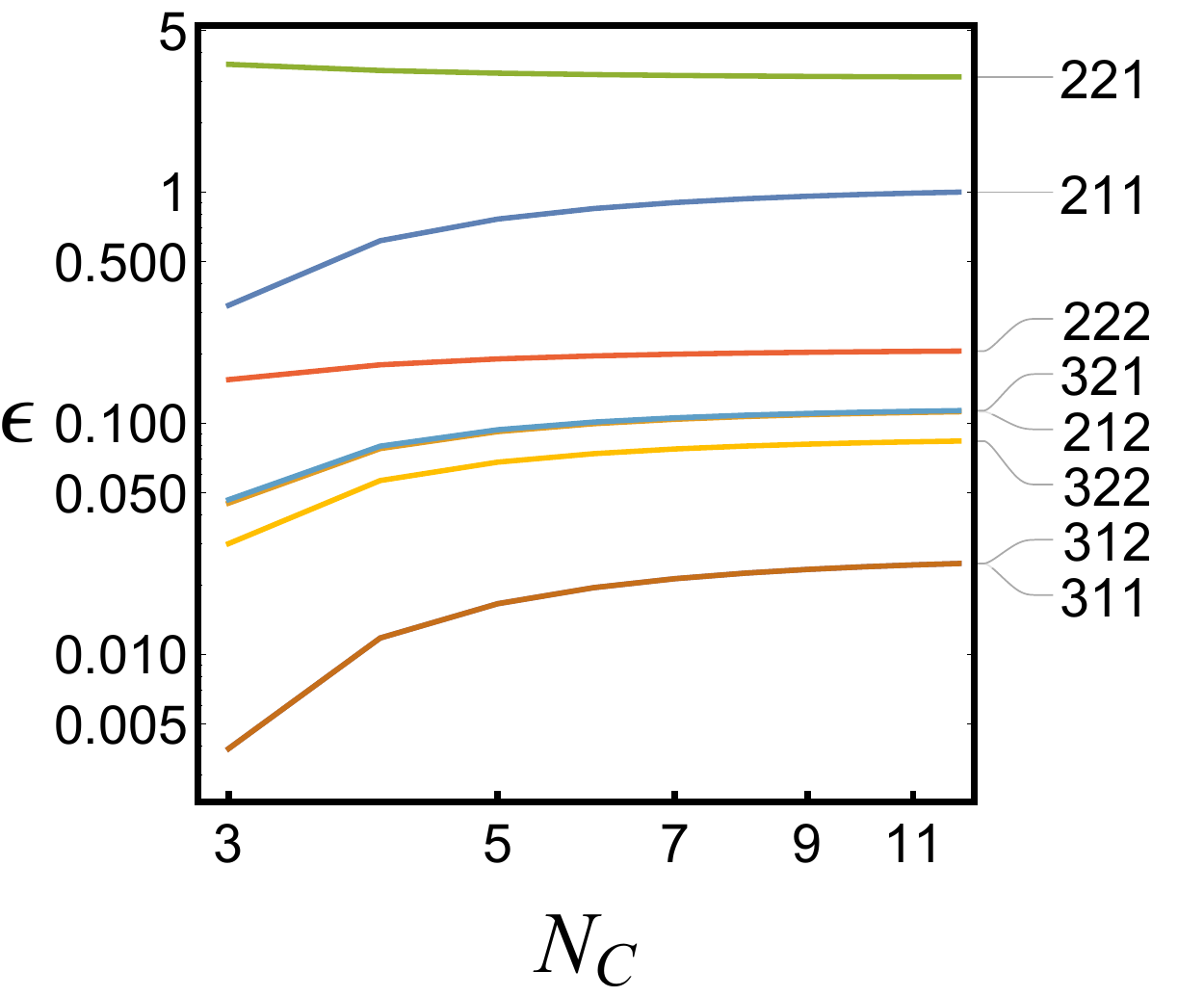}
\caption{Upper boundary for the parameter $\epsilon$ for various approximations as functions of $\nc$ (see Tab.~\ref{tbl:eps}).}
\label{fig:eps_max_all}
\end{figure}

\begin{figure}[b]
    \includegraphics[width=.8\linewidth]{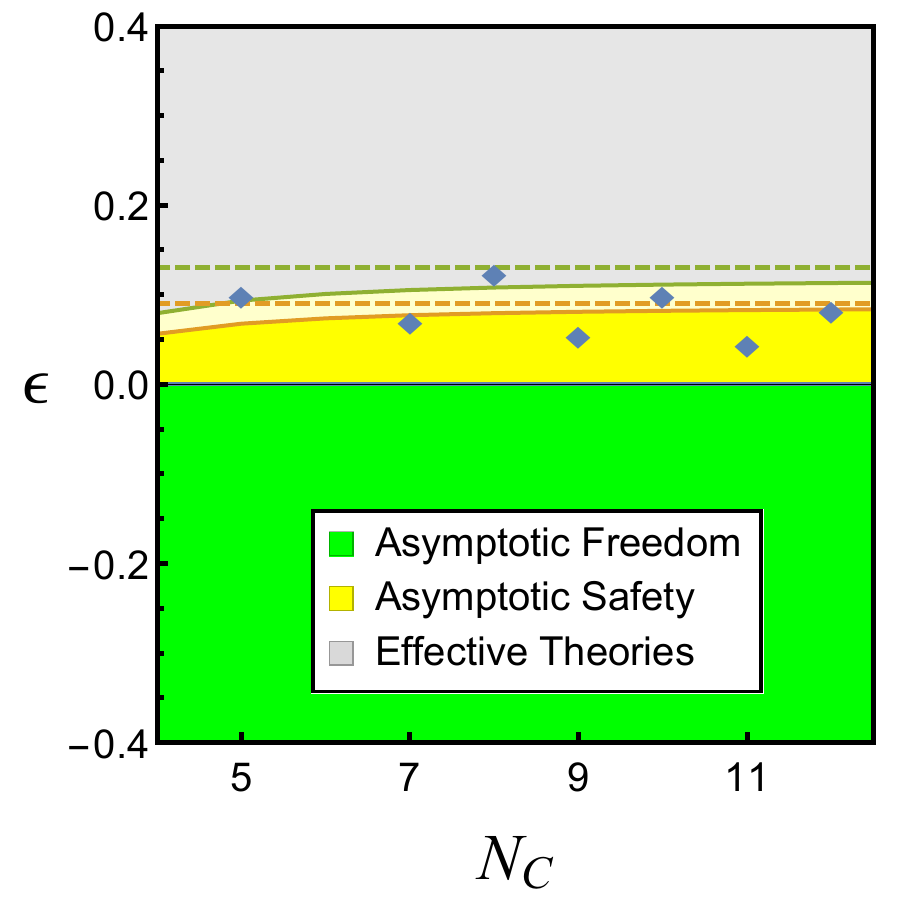}
    \caption{Conformal window from beta functions up to NNLO as given by $\epsilon_{\rm subl.}$ comparing approximations {\texttt{321}} and {\texttt{322}}. Dashed lines represent the asymptotic value and dots the first integer solutions of $\epsilon$.}
    \label{fig:cw2}
\end{figure}

In order to obtain a better understanding of the system we have not restrained ourselves to consistent approximations only, but have also computed all other possible combinations with the beta functions that we have available. The first key result of this section is in Table \ref{tbl:eps}, showing the boundaries of the conformal window in the $\epsilon_{subl.}$ approach. Next, we give several comments to explore the significance of this table. First, we reproduce and compare the results in the Veneziano limit with the cases $\nc=\{5,7\}$, showing that the overall trend is to narrow the conformal window as $\nc$ becomes smaller (with the exception of {\texttt{221}}). This continues to reflect the nature of the $f(\nc)$ functions introduced in the previous section, where we saw that they drive $\epsilon_{max}$ to smaller values at lower $\nc$. Second, note that the comparative relations accross approximations are mantained for all the three cases shown (e.g. {\texttt{311}} is always more strict than {\texttt{211}}), meaning that no particular approximation is favoured at low $\nc$. Third, at each approximation the conformal window remains constrained by the same source in all three cases (a. strong coupling, b. fixed point merger, c. vacuum instability). Thus, finite $\nc$ corrections do not qualitatively change the behaviour of the system of beta functions.

\begin{table}[t]
\aboverulesep = 0mm
\belowrulesep = 0mm
	\addtolength{\tabcolsep}{1pt}
	\setlength{\extrarowheight}{1pt}
\begin{tabular}{`cccc`}
 \toprule
  \rowcolor{Yellow}
{\bf $\nc$} &\bf $\epsilon_{\rm max}$ & \bf $N_{\rm Fmin}$ & \bf $N_{\rm Fmax}$\\
\midrule
 $\nc=\infty$ &0.087\, \,  & $\s0{11}2\nc$ & $(\s0{11}2+\eps_{\rm max})\nc$ \\
\rowcolor{LightGray}
 $\nc=9$ & 0.0810 & 50 & 50 \\
 $\nc=7$ & 0.0771 & 39 & 39 \\
\rowcolor{LightGray}
 $\nc=5$ & 0.0677 & -- & -- \\
 $\nc=3$ & 0.0300 & -- & -- \\
\bottomrule
\end{tabular}
\caption{Bounds based on beta functions, and selected integer value solutions $(\nc,\nf)$ within the UV conformal window of Fig.~\ref{fig:cw2}.
The hyphen indicates that no integer solution for $N_{\rm f}$ can be found.}
\label{tbl:valuesNC2}
\end{table}

The second key result is Figure \ref{fig:eps_max_all}, where a plot of the boundaries is presented for extended values of $\nc$. The first striking feature in this plot is that it clearly shows how $\epsilon_{\rm subl.}$ quickly converges to the asymptotic value (e.g. for {\texttt{322}}, at $\nc=7$, the value differs from the asymptotic value by only 10.9\%).  Furthermore, we observe that all approximations share roughly the same rate of convergence. Second, we can clearly visualize and confirm the qualitative picture obtained from \eqref{eq:betas_shift_211} and \eqref{eq:betas_shift_322}. To highlight this, we note that approximation {\texttt{221}} is the least constrained, which is in line with our prediction that the running of the Yukawa interaction widens the conformal window. On the other hand, we have {\texttt{311}} and {\texttt{312}}, clearly showing that the running of the gauge coupling leads to tighter constrains. Overall, we conclude that finite $N$ corrections shrink the conformal window.

\subsection{UV Conformal Window Revisited}

Finally, we revisit estimates for the conformal window using the data for $\epsilon_{\rm subl.}$ in Figure \ref{fig:cw2}. In the plot we include approximations {\texttt{321}} and {\texttt{322}} as the light and dark yellow areas respectively. For comparison we have also superimposed the asymptotic value of $\epsilon_{\rm subl.}$ as dashed lines of the corresponding color, and dots representing the smallest values of $\epsilon$ for integer number of fields. The tightest constraint at {\texttt{321}} and {\texttt{322}} arises due to a fixed point merger and the onset of vacuum instability, respectively.

\begin{figure}[t]
\centering
\includegraphics[width=.9\linewidth]{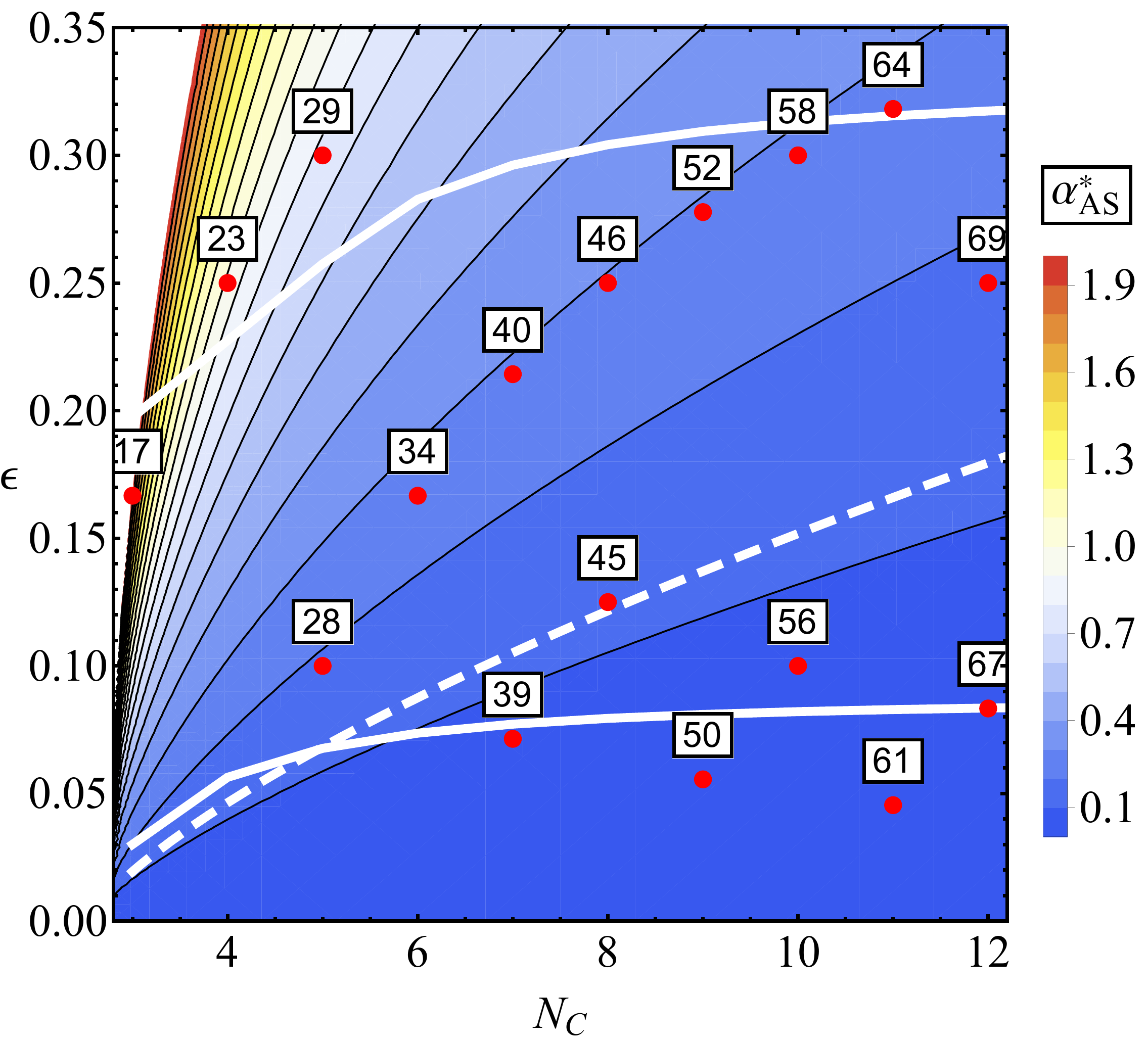}
\caption{Contour plot of ultraviolet fixed points 
in the $(\nc,\eps)$ plane, color-coded according to the magnitude of  $\alpha_{\rm AS}^*$. The white dashed line represents the value of the QCD coupling at the Z boson pole mass $\alpha_s(M_Z^2)$.
Red dots indicate fixed points for integer $(\nc,\nf)$ with $\nf$ given in the square box. The upper (lower)  full white curve indicates the  bound of the conformal window   from fixed points (beta functions).}
\label{fig:alphaAS2}
\end{figure}

Although the exact boundary of the conformal window narrows at small $\nc$, we recall that the field multiplicities can only be integer numbers, and paying close attention, we can notice that the $\nc$ dependence does not exclude any integer solution that was not excluded already in the Veneziano limit (all the blue dots below the dashed line, are also below the solid line). The first few integer solutions are shown in Tab.~\ref{tbl:valuesNC2}. We expect that the ranges indicated in Fig.~\ref{fig:cw1} (and also Fig.~\ref{fig:alphaAS2}),  account well for the uncertainties in the perturbative estimate of the lower bound (see Fig.~\ref{fig:eps_max_all}).

\subsection{Perturbativity and Comparison with QCD}
Away from the strict Veneziano limit, perturbativity of the fixed point is not automatically guaranteed and a relevant question is if any of the theories admitted in the conformal window fall within the perturbative domain. Although there is not an strict boundary on what is perturbative and what is not, we address this question by comparing with perturbative QCD. Specifically, at the mass of the Z boson, where perturbation theory is applicable, the  running coupling constant  of QCD has been measured to be \cite{Bethke:2006ac}
\begin{align}
\alpha_s\left(M_Z^2\right) &= 0.1185 \pm 0.0006\,.
\end{align}
In order to make a sensible comparison, we normalise the asymptotically safe gauge coupling in exactly the same manner, writing
\begin{align}
\label{eq:ag_norm}
\alpha_{\rm AS}^* &= \frac{4\pi \alpha_g^*}{\nc}\,,
\end{align}
with $\alpha_g^*$  given by \eqref{eq:ag_nc} to order $O\left(\epsilon^2\right)$. 

Next, we compare our results for fixed point values in various asymptotically safe models with the value of the QCD coupling constant at the mass of the Z boson. Fig.~\ref{fig:alphaAS2}  shows a contour plot of ultraviolet fixed points $\alpha_{\rm AS}^*$ in the $(\nc,\eps)$ plane. Red dots indicate fixed points with integer $(\nc,\nf)$ with $\nf$ indicated in the box.  The upper (lower)  full white curve indicates the  conformal window   from fixed points (beta functions). The white dashed line represents the QCD coupling at the pole mass of the Z boson $\alpha_s(M_Z^2)$.

Theories falling within the first shaded region (counting from bottom to top) have a fixed point gauge coupling smaller than $\alpha_s\left(M_Z^2\right)$, and can be considered ``more perturbative" than perturbative QCD. In the second shaded region the value of the coupling is between 0.1 and 0.2; for the next one 0.2 and 0.3, and so on. We conclude that  it should be possible to identify asymptotically safe quantum field theories even at finite $\nc$ and beyond the Veneziano limit using perturbation theory.

\begin{figure}
	\includegraphics[width=.8\linewidth]{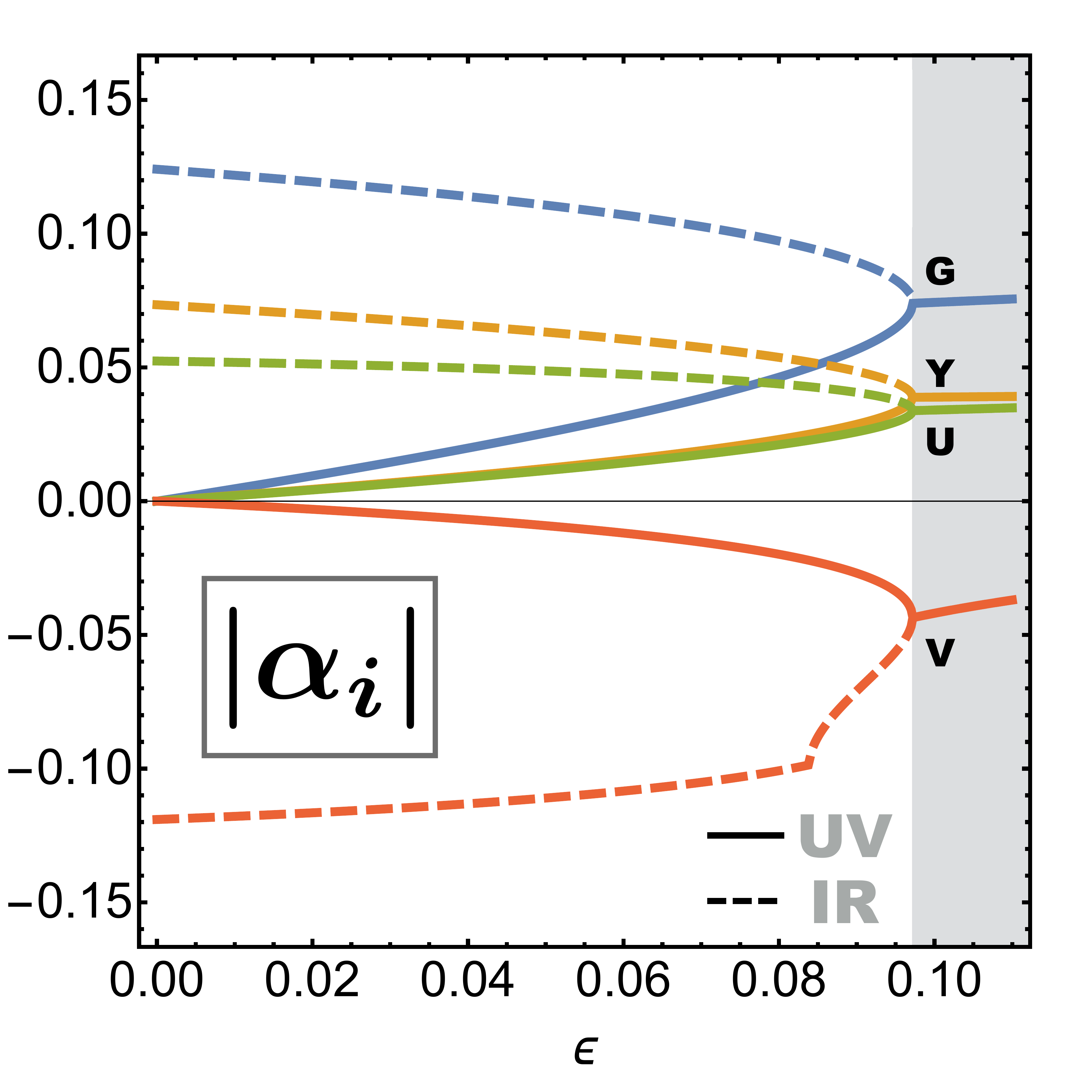}
		\caption{Coupling values (real part) of the UV (solid lines) and IR (dashed lines) fixed points as a function of $\epsilon$ in the Veneziano limit at approximation  order \texttt{322}. From top to bottom, the lines correspond to the gauge, Yukawa, single trace scalar and double trace scalar couplings.}
	\label{fig:couplings_irfp}
\end{figure}

\section{\bf Interacting IR Fixed Points}\label{IR}
It is well-known that the theories \eq{L}  display  infrared Banks-Zaks  fixed points in the regime with asymptotic freedom  \cite{Litim:2014uca,Bond:2018oco,Bond:2016dvk}, and for sufficiently small $|\eps|\ll 1$. 
In this section, we point out that the theory \eq{L} also displays interacting infrared fixed points in the regime where asymptotic freedom is absent. We then discuss whether the new IR fixed point is responsible for the disappearance of the interacting UV fixed point through a merger.

\subsection{Colliding Fixed Points}
We consider the regime $\eps>0$ where asymptotic freedom is absent. As discussed in the previous sections, the theory \eq{L}  displays a primary interacting UV fixed point whose couplings  \eq{eq:ag_nc} and scaling exponents \eq{eq:eig1} arise as strict power series in the small parameter $\eps$. Interestingly though, 
a secondary fixed point $\alpha^*_{\rm IR}$ can arise starting 
from the 3-loop order in the gauge coupling,   
and can be expressed as a power series in $\epsilon$ with a leading constant term. Such a constant term indicates that the solution is not  under rigorous perturbative control even for small $\eps$, and must be treated with care.

\begin{figure}[t]
	\includegraphics[width=.8\linewidth]{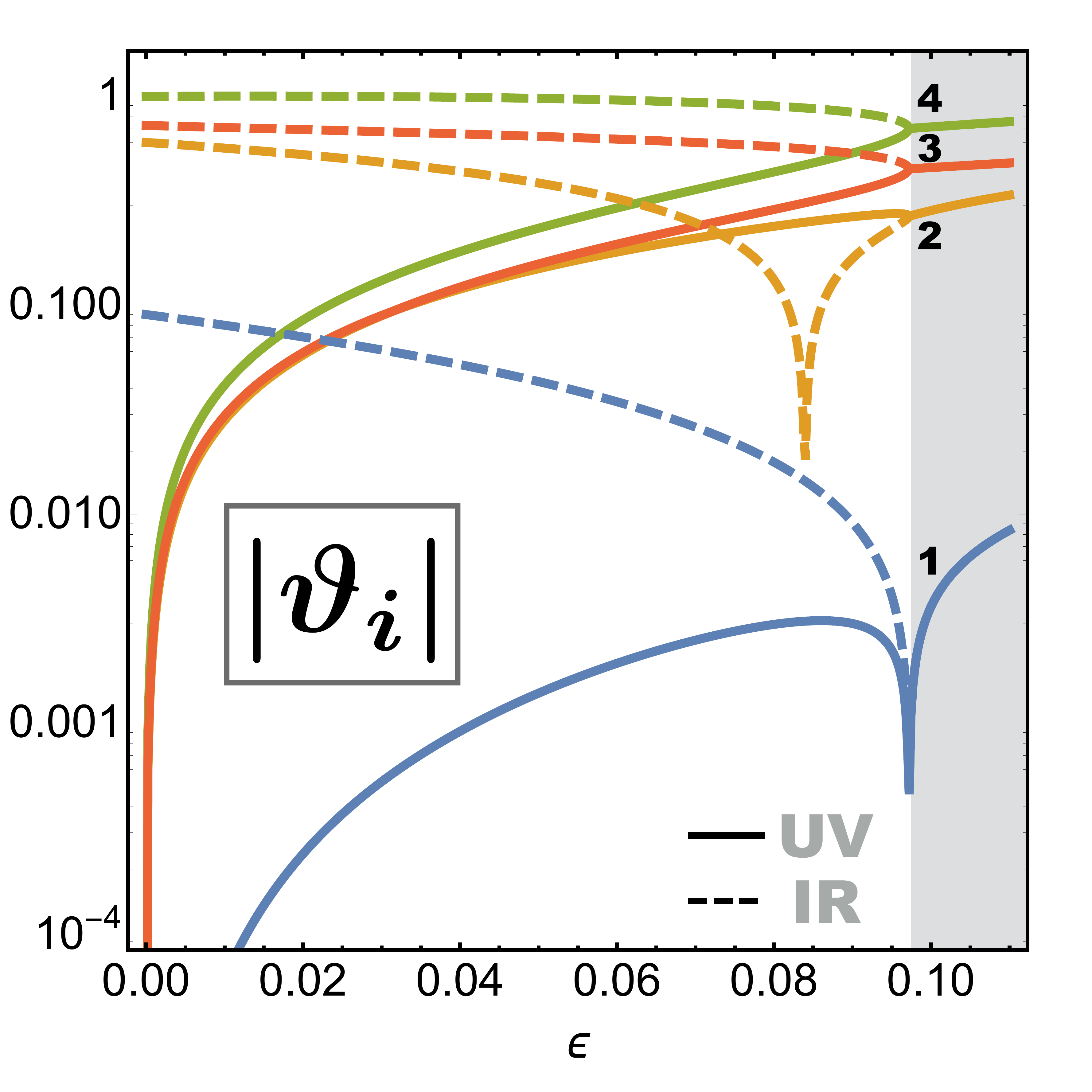}
	\caption{Eigenvalues (absolute value) of the UV (solid lines) and IR (dashed lines) fixed points as a function of $\epsilon$ in the Veneziano limit at approximation order  \texttt{322}. From top to bottom, the lines correspond to eigenvalues dominated by the single trace scalar, double trace scalar, Yukawa and gauge couplings.}
	\label{fig:eigenvalues_irfp}
\end{figure}

With this disclaimer in mind, we start by displaying  our results  
in Fig.~\ref{fig:couplings_irfp}. It shows the fixed point couplings, exemplarily in the \texttt{322} approximation and  in the Veneziano limit, with full (dashed) lines referring to the UV (IR) fixed point.  We observe that the secondary fixed point $\alpha^*_{\rm IR}$ has no UV attractive directions, and therefore takes the role of an IR sink. Its fixed point couplings are slightly larger in magnitude than those at the corresponding UV fixed point \eq{eq:ag_nc}, decreasing in magnitude with increasing $\epsilon$. Most notably, as $\epsilon$ grows, the UV and IR fixed points get closer to each other until they meet and annihilate at  $\eps=\epsilon_{\rm merge}$.  The merging of fixed points is one of the fundamental mechanisms by which physical fixed points can disappear into the complex plane \cite{Miransky:1996pd,Kaplan:2009kr,Braun:2010qs,Gukov:2016tnp,Kuipers:2018lux}, indicated by the  gray-shaded area in Fig.~\ref{fig:couplings_irfp}.

In  Tab.~\ref{tbl:eps_merger} we compute $\eps_{\rm merge}$ for different additional approximations in the scalar and Yukawa sectors, comparing the \texttt{311}, \texttt{321}, and \texttt{322} approximations. We observe that the value of $\eps$ at the merger depends on the approximation. Moreover, for decreasing $N_{\rm c}$, we also find that $\eps_{\rm merge}$ decreases. Comparing the  \texttt{321} and \texttt{322} approximations, we note that the 2-loop quartic contributions only have a mild impact. In turn, comparing the  \texttt{311} with the \texttt{321} and \texttt{322} approximations, we note that the 2-loop Yukawa contributions are quantitatively important. This is  consistent with the analogous analysis for the UV fixed point, summarised in Tab.~\ref{tbl:eps}.
 We stress that  $\epsilon_{\rm merge}$ is numerically small in all cases, and we  expect perturbation theory to remain viable at least in some vicinity around $\eps_{\rm merge}$.

\begin{table}[b]
\aboverulesep = 0mm
\belowrulesep = 0mm
	\addtolength{\tabcolsep}{1pt}
	\setlength{\extrarowheight}{1pt}
\begin{tabular}{`cccc`}
	\toprule
	\rowcolor{Yellow} \multicolumn{4}{`c`}{Critical endpoint $\bm{\epsilon_{\rm merge}}$} \\
	\midrule
	$N_{\rm c}$  & \texttt{311} & \texttt{321} & \texttt{322} \\
	\midrule
\rowcolor{LightGray}		$\infty$ & 0.02653 & 0.1170 & 0.09717 \\
10	 & 0.02387 & 0.1113 & 0.09161 \\
\rowcolor{LightGray}		7	 & 0.02121 & 0.1053 & 0.08574 \\
	5	 & 0.01654 & 0.09350 & 0.07442 \\
\rowcolor{LightGray}		3	 & 0.003906 & 0.04659 & 0.03096 \\
	\bottomrule
\end{tabular}
\caption{Value of $\epsilon_{\rm merge}$ for various approximation orders in the loop expansion. The trend is that the merger occurs at higher values of $\epsilon$ as $N_{\rm c}$ increases up to a maximum in the Veneziano limit.}
\label{tbl:eps_merger}
\end{table}

Fig.~\ref{fig:eigenvalues_irfp} shows the absolute value of scaling exponents at the UV and IR fixed points as functions of $\eps$. Note that this is a logarithmic plot where vanishing eigenvalues correspond to a sharp downward peak. 
We observe that the scaling exponents of both fixed points coincide at the merger point.  In the limit $\eps\to \epsilon_{\rm merge}$, the sole relevant eigenvalue $\vartheta_1$ of the UV fixed point and one of the IR eigenvalues both become exactly marginal (blue curves), which again is a characteristic fingerprint of a fixed point merger. Fixed points become complex in the grey-shaded area. Miransky scaling \cite{Miransky:1996pd}, which is a well-known feature of fixed point mergers \cite{Kaplan:2009kr,Braun:2010qs,Gukov:2016tnp,Kuipers:2018lux}, is also observed at the endpoint.

\begin{figure}
		\includegraphics[width=.8\linewidth]{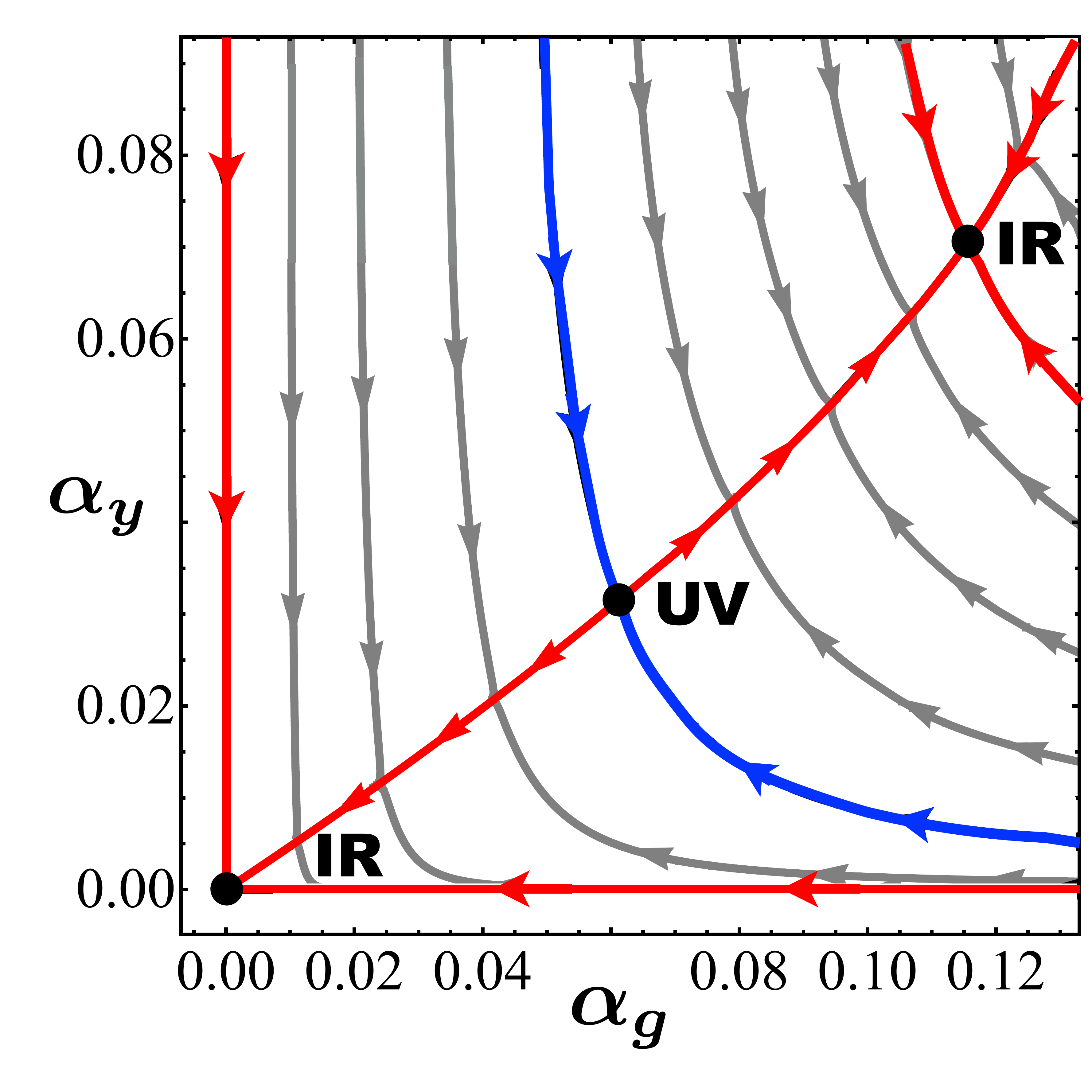}
	\caption{Phase diagram  in the $(\alpha_g,\alpha_y)$ plane 
	close to the merger limit, showing the interacting UV and IR fixed points,  the free IR fixed point, and  sample RG trajectories. Quartic couplings are projected onto the plane with $\beta_u=\beta_v=0$.}
	\label{fig:PD}
\end{figure}

\begin{figure*}
	\includegraphics[width=.7\linewidth]{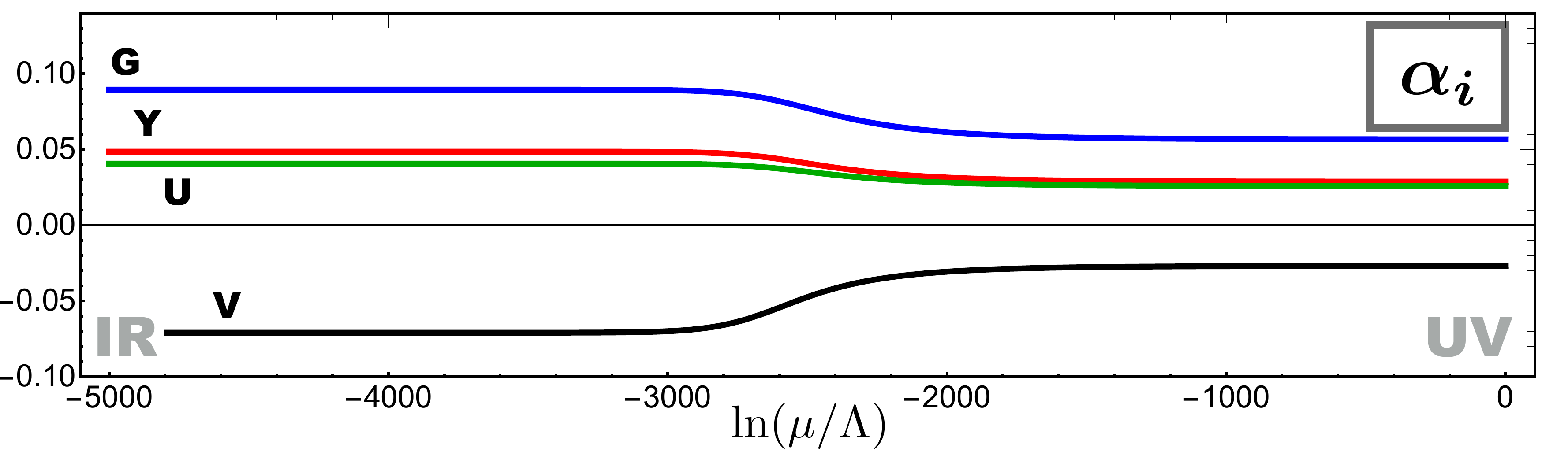}
	\caption{Renormalization group  trajectory connecting the interacting UV and IR  fixed points in the Veneziano limit. From top to bottom, the lines $\alpha_i$ correspond to the gauge, Yukawa, single, double trace scalar couplings.}
	\label{fig:rgflow_irfp}
\end{figure*}

\subsection{IR Conformal Window}
Next, we discuss the conformal window associated to the new IR fixed point $\alpha^*_{\rm IR}$. Since couplings are of the order of $\eps_{\rm merge}$ and thus 
small close to the fixed point merger, we may view $(\eps_{\rm merge}-\eps)$ 
as a small parameter to study the IR conformal window. 
Using \eq{eq:beta_g}, \eq{eq:beta_y}, \eq{eq:beta_u} and \eq{eq:beta_v} in the \texttt{311} and  the \texttt{321} approximations, we then find that the IR conformal window
covers the range $\eps\in (0,\eps_{\rm merge})$. On the other hand,
in the \texttt{322} approximation,  
the lower limit of the conformal window arises through another  merger, at about $\eps_{\rm IR}\approx 0.085$, and hence $\epsilon\in(0.085,0.90)$ approximately.
This can be appreciated in Fig.~\ref{fig:eigenvalues_irfp}, where the downward peak in $\vartheta_2$ around $\epsilon\approx0.085$ announces another fixed point merger (the tertiary fixed point responsible for this merger is not shown). In Fig.~\ref{fig:couplings_irfp}, the secondary merger
is visible in the coupling $\alpha_v$, which becomes complex. In the Veneziano limit, 
$\beta_v$ decouples from the system of equations. 
 Therefore, $\alpha_v$ may take complex values without disturbing the other couplings. For finite $N_{\rm c}$, however, all couplings become complex  at the merger. In either case, we conclude that a reliable lower bound on the value of $\epsilon$ cannot be found within perturbation theory. 
We further note that  the scaling exponents associated to the gauge and Yukawa couplings become complex conjugate for small values of $\epsilon$ and $N_{\rm c}\geq5$ in the \texttt{321} approximation. This is however not reproduced in any of the other approximations. We interpret both of these effects as shortcomings of the approximations, and as a sign that the IR fixed point may only be trusted  close to $\epsilon_{\rm merge}$. Future work using higher orders in perturbation theory, or non-perturbative continuum and lattice methods, are required to confirm the existence of the IR fixed point $\alpha^*_{\rm IR}$ for small $\eps$, away from the merger.

\subsection{Phase Diagram}

For $\eps$ close to but below $\epsilon_{\rm merge}$, both the UV and IR fixed point are under reasonable perturbative  control,
and we may therefore ask how the corresponding phase diagram  looks like. This is shown in Fig.~\ref{fig:PD} using exemplarily 
the \texttt{321} approximation with $\eps=\s01{10}$ (similar phase diagrams are found within the  \texttt{311} and  \texttt{322}  
approximations). We observe the Gaussian IR fixed point, and the interacting UV and IR fixed points.  The irrelevant directions
at the UV fixed point are shown in blue, and separatrices are shown in red with arrows pointing from the UV to the IR. 
As an aside, we note that a phase diagram such as Fig.~\ref{fig:PD} cannot  arise for a non-asymptotically free supersymmetric QFT.\footnote{The  main
reason for this is that 
both fixed points are fully interacting. With $N=1$ supersymmetry, the additional global and anomaly-free $U(1)_R$ symmetry   
then implies identical $R$-charges  and identical values for the central charge $a$. Consequently, in supersymmetry, 
a flow from one interacting fixed point to the other would be in conflict with the $a$ theorem. 
Still, asymptotically safe UV fixed points can arise in certain semi-simple supersymmetric QFTs \cite{Bond:2017suy}.}
As such, the phase diagram Fig.~\ref{fig:PD} should be viewed as a  feature of non-asymptotically free and 
non-supersymmetric quantum field 
theories.

Fig.~\ref{fig:rgflow_irfp} shows the running couplings along the separatrix connecting the interacting UV and IR fixed points using the \texttt{322}  
approximation with $\eps=0.09$. 
Couplings show a smooth cross-over from conformal scaling in the UV to conformal scaling in the IR.
As such, our model is an example of a non-supersymmetric theory which displays non-trivial conformal fixed points both in the UV and in the IR.
We may further conclude that the theories converge towards interacting and unitary four-dimensional conformal field theories in the asymptotic high- and low-energy limits, as has been explained in \cite{Luty:2012ww} using techniques related to the proof of the $a$-theorem \cite{Komargodski:2011vj,Komargodski:2011xv}.

\section{\bf Matter-Dominated Fixed Points} \label{Strong}
\label{sec:strong_coupling}
The results of the previous sections arise in the regime where 
approximations are under perturbative control ($\eps\ll 1$). 
This is also the regime where  fixed point are the result of 
a balance  between matter and gauge field fluctuations, with $\nf/\nc$ of the order of a few.
In this section, we discuss the prospect for fixed points in the regime where 
matter field fluctuations dominate over those by the gauge fields. Parametrically,
this corresponds to  taking $\nf$ much larger than $\nc$  ($\epsilon \gg 1$). 

This idea has  initially been looked into  in  the infinite $\nf$ limit with $\nc$ kept fixed,  both for abelian  \cite{PalanquesMestre:1983zy} and  non-abelian gauge theories
 \cite{Gracey:1996he} by using  all-order resummations  in the $\overline{\rm MS}$ scheme and critical-point methods, respectively.
If applicable, results then suggest  the existence of an ultraviolet Banks-Zaks fixed point owing to a negative singularity of the resummed beta function. Further work on the possible existence of such a fixed point  \cite{Holdom:2010qs,Shrock:2013cca,Litim:2014uca,Antipin:2017ebo,Kowalska:2017pkt,Alanne:2018csn,Dondi:2019ivp,Alanne:2019meg,Sannino:2019vuf,Dondi:2020qfj} and proposals for BSM models incorporating such a solution have already been put forward, e.g.~\cite{Abel:2017rwl,Pelaggi:2017abg}. 

On the other hand, it has also been pointed out that the negative singularity, and hence the fixed point, might be an artifact of the large $\nf$ limit.  Several indicators are pointing into this direction:
\begin{itemize}
\item In QED, the fermion mass anomalous dimension diverges at the fixed point \cite{PalanquesMestre:1983zy}. 
\item In $N=1$ supersymmetric gauge theories  coupled to matter, infinite order resummed beta functions 
are scheme dependent   \cite{Martin:2000cr}: a negative singularity and a fixed point do arise  in the Novikov-Shifman-Vainstein-Zhakarov scheme, but not so in the DRED scheme, leading to the conclusion that the infinite $\nf$ fixed point cannot be trusted \cite{Martin:2000cr}.
\item In QCD the glueball anomalous dimension violates the unitarity bound close to the fixed point \cite{Ryttov:2019aux}.  
\item A more general analysis of higher-order  corrections  came to the conclusion  that the putative 
fixed point no longer arises at finite  $\nf$   \cite{Alanne:2019vuk}. 
\item Advanced lattice simulations, although not conclusive,  have  not found any support for this type of fixed point in QED \cite{Leino:2019qwk}.  
\end{itemize}
For these reasons, there is presently not sufficient evidence for the validity of a large $\nf$ fixed point  in the gauge sector. 

In this light, our take on this  will be through the role of Yukawa and scalar self-interactions. At weak coupling, it is well-known that this allows for new scaling limits, qualitatively different from Banks-Zaks fixed points. Here, we clarify whether Yukawa interactions may also provide a new  scaling limit at finite or infinite $\nf\gg \nc$,  different from the one seen in the gauge sector. 

\subsection{Banks Zaks at Large $\eps$} 
\label{sub:finite_order_perturbation_theory}

To set the stage, we first recall the setup of \cite{PalanquesMestre:1983zy, Gracey:1996he, Holdom:2010qs} and consider a gauge theory with fermions transforming under the representation $R$. In perturbation theory, the \textit{p}-loop order term is a polynomial in $\nf$ of degree \textit{p-1}. The highest order term in $\nf$ corresponds to diagrams with internal fermion propagator bubbles. In the large $\nf$ limit, these diagrams with internal chains of bubbles dominate the renormalisation group flow. By rescaling the coupling $\alpha \to S_R \nf \alpha/\pi$, where $S_R$ is the Dynkin index of the fermions' representation, one can reorganize the beta function from a power expansion in small coupling to a power expansion in $1/\nf$. Through the use of a clever resummation technique, an exact all-orders expression for the beta function has been obtained for abelian \cite{PalanquesMestre:1983zy} and non-abelian gauge groups \cite{Gracey:1996he} (see also \cite{Holdom:2010qs}). Keeping the notation used in this study it can be written as
\begin{align}
\beta_g = \frac{2 A}{3}\, \alpha_g\,\left(1 + \sum_{i=1}^\infty \frac{H_i(A)}{\nf^i}\right)
\end{align}
with $A = 4 \nf S_R \alpha_g / \nc$ and $H_I(A)$ a coefficient encoding the contribution of diagrams with the highest powers of $\nf$ at all orders. This expression is arranged such that the leading term of the sum is $2A \alpha_g/3$, corresponding to the one loop result. The next to leading term is
\begin{align}
\begin{split}
\textstyle H_1(A) &= \textstyle -\frac{11}{4} \frac{C_G}{S_R} + \int_0^{A/3}dx\, I_1(x) I_2(x)  \\
\textstyle I_1(x) &= \textstyle \frac{(1 +x)(2x -1)^2 (2x -3)^2 \text{sin}(\pi x)^3 \Gamma(x-1)^2 \Gamma(-2x)}{(x-2) \pi^3}\\
\textstyle I_2(x) &= \textstyle \frac{C_R}{S_R} + \frac{20 -43x +32x^2 -14x^3 +4x^4}{4(2x-1)(2x-3)(1-x^2)} \frac{C_G}{S_R}
\end{split}
\end{align}
The case for a non-trivial fixed point comes from the poles in the terms $I_1$ and $I_2$. The divergence drives $H_1$ towards large negative values such that the $1/\nf$ suppression is bested and a cancellation with the one loop term is possible. The pole occurs at  $A_* = 15/2$ for the abelian case and at $A_* = 3$ for the non-abelian case, which would imply a fixed point  $\alpha_g^*=A_* \nc/( 4 \nf S_R)$ with parametrically large scaling exponent $1/\vartheta \to 0$.

Next, we consider finite loop order approximations. 
The integral in $H_1$ can be evaluated by first performing a Taylor expansion of the integrand around vanishing \textit{x}. Keeping the first two terms of the expansion one gets
\begin{align}
\begin{split}
\beta_g & = 
-\frac{4}{3}\left( \frac{11}{2}C_2^G - 2\nf S_2^R \right) \frac{\alpha_g^2}{\nc} \\
&\mathrel{\phantom{=}} +\left( \frac{40}{3} C_2^G + 8 C_2^R \right)\nf S_2^R \frac{\alpha_g^3}{\nc^2} \\
&\mathrel{\phantom{=}} -\frac{1}{27}\left( 79 C_2^G + 66 C_2^R\right) \left(2 \nf S_2^R\right)^2 \frac{\alpha_g^4}{\nc^3}
\end{split}
\end{align}
which precisely corresponds to the terms with highest power of $\nf$ at the first three loop orders in the perturbative beta function, see \eqref{eq:beta_g1_R}-\eqref{eq:beta_g3_R} in the Appendix.
Then, at finite loop order, the fixed point arises through a cancellation  between the first and the highest loop order available. In our notation, large $\nf$ corresponds to $1/\epsilon \to 0$, so that at three loop and omitting factors of $\nc$, we have
\begin{align}
\beta_g^{(1)} \propto \epsilon\, \alpha_g^2,\quad \beta_g^{(2)} \propto \epsilon\, \alpha_g^3,\quad \beta_g^{(3)} \propto \epsilon^2\, \alpha_g^4
\end{align}
For a cancellation to occur between the first and the third loop order, $\alpha_g^*$ must scale as $\alpha_g^* \sim 1/\sqrt{\epsilon}$. With this scaling, the second loop term becomes subleading in $\epsilon$, so the cancellation is purely between first and third loop order. At the \textit{n}th loop order, the scaling of the coupling is $\alpha_g^* \sim \epsilon^{(2-n)/(n-1)}$ (provided that the \textit{n} loop order is overall negative) \cite{Pica:2010xq,Shrock:2013cca}. In the absence of scalar fields, the gauge coupling takes the fixed point value,
\begin{align}\label{alphaBZ} 
\begin{split}
\alpha_g^*\big|_{\texttt{300}}  &= \frac{3}{2\sqrt{7}} \frac{\nc}{\sqrt{\nc^2 -\frac{33}{112}}} \frac{1}{\sqrt{\epsilon}} \\
&\mathrel{\phantom{=}} +\frac{117}{112} \frac{\nc^2 -\frac{3}{13}}{\nc^2 -\frac{33}{112}} \frac{1}{\epsilon} 
 + O\left(\frac{1}{\epsilon^{3/2}}\right)\,,
\end{split}
\end{align}
and taking $\nc \to \infty$ we recover the result of \cite{Litim:2014uca}. This is the behaviour expected
from the infinite-$\nf$ resummed result.
We also reproduce the scaling exponent
\beq\label{thetaBZ}
    \vartheta_1\big|_{\texttt{300}} = 
		-\frac{4}{\sqrt{7}}\frac{\nc}{\sqrt{\nc^2 -\frac{33}{112}}}\sqrt{\epsilon} 
		+O\left(\frac{1}{\epsilon^{1/2}}\right)\,.
\eeq
Taking the result at face value, we find a lower bound for $\eps$
due to the onset of large coupling values,
\begin{align}\nonumber
\begin{split}
    \epsilon_{\rm min}
    =& 
\textstyle			\frac{477
		\left(\nc^4 +\frac{55}{318}\nc^2 -\frac{9}{318}\right)}{224\nc^2\left(\nc^2 -\frac{33}{112}\right)}
	\textstyle		+\frac{57\sqrt{505}}{224\nc^2\left(\nc^2-\frac{33}{112}\right)} \times
\\&
\textstyle				\sqrt{\nc^8 -\frac{12925}{36461}\nc^6 
		+\frac{14461}{729220}\nc^4 +\frac{1683}{364610}\nc^2+\frac{81}{729220} }\,.\end{split}
\end{align}
In previous works  \cite{Pica:2010xq,Shrock:2013cca,Litim:2014uca}, the expressions \eq{alphaBZ}, \eq{thetaBZ} have been interpreted as the finite loop order remnant of the infinite order fixed point. 
On the other hand, given the before-mentioned points of critique which indicate 
that the zero at infinite order are an artifact of the large-$\nf$ limit, we conclude 
that the finite order image of the Banks-Zaks fixed point candidate at large $\eps$ cannot be trusted either. 
It is then interesting to check whether Yukawa and quartic couplings  can modify the outcome, to which we turn next.

\begin{figure}[t]
\centering
\includegraphics[width=.8\linewidth]{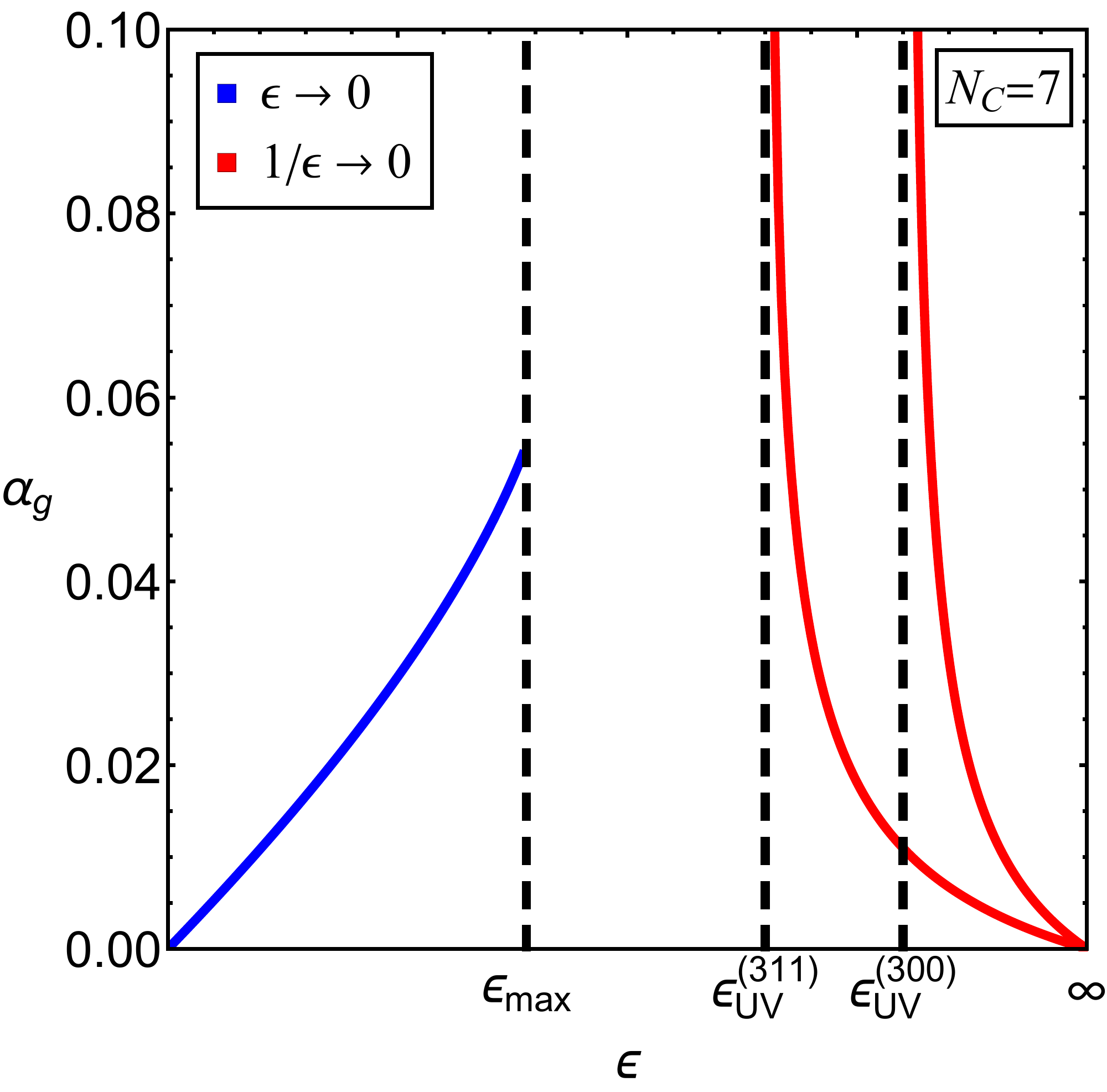}
\caption{Comparison of the conformal window at weak  coupling ($\epsilon\ll 1$, blue) with  conjectured ones at strong coupling ($\epsilon\gg 1$, red), evaluated at $\nc=7$.
For small $\eps$, the couplings at the GY fixed point remain perturbative throughout. For large $\eps$, the BZ (\texttt{300} approximation) and the GY (\texttt{311} approximation)  fixed point candidates coincide asymptotically, and deviate mildly when approaching their lower bounds  where $\alpha_g$ diverges (see main text).
}
\label{fig:cw_sl}
\end{figure}

\subsection{Gauge-Yukawa  at Large $\eps$} 
\label{sub:yukawa_coupling at leading order}
Given the above state of affairs, we now include effects from uncharged scalar fields and Yukawa interactions. 
For the model at hand,  no infinite order in $N_{\rm f}$  expressions for beta functions are available. 
Therefore, we resort to finite order perturbation theory and mimick the analysis of the previous section. 

We begin with retaining the Yukawa and scalar interactions to first loop order and in the  $1/\epsilon \to 0$ limit. The corresponding beta functions read
\begin{align}
\begin{split}
\beta_y^{(1)} &\approx 2\epsilon\, \alpha_y^2 - 6 \alpha_g \alpha_y\\
\beta_u^{(1)} &\approx 8\alpha_u^2 + 4\alpha_y \alpha_u - 2\epsilon\, \alpha_y^2 + 48 \nc^{-2} \epsilon^{-2}\, \alpha_u \alpha_v\\	
\beta_v^{(1)} &\approx 12\alpha_u^2 + 4\alpha_y \alpha_v + 4\alpha_v^2 + 16\alpha_u \alpha_v\,.
\end{split}
\end{align}
For the Yukawa beta function to vanish, the cancellation must occur within the same loop order. This implies that at the fixed point, the Yukawa coupling must scale as: $\alpha_y^* \sim \alpha_g^*/\epsilon \sim 1/\epsilon^{3/2}$. To find the scaling of the scalar interaction $\alpha_u$, we note that the cancellation should not depend on the $\alpha_u \alpha_v$ term, as this vanishes for large $\nc$. 
Next, we rule out that the cancellation could depend on $\alpha_y \alpha_u$ noting that the implied scaling would make $\alpha_u^2$ more relevant and the beta function would not vanish. The only possible choice is for the negative term to cancel with $\alpha_u^2$, implying that at the fixed point $\alpha_u \sim \sqrt{\epsilon}\, \alpha_y \sim 1/\epsilon$. A similar argument follows for the double trace scalar self interaction, finding the same scaling $\alpha_v \sim \sqrt{\epsilon}\, \alpha_y \sim 1/\epsilon$. These interactions modify the fixed point of the gauge coupling only at subleading order. Overall, we therefore find
\begin{widetext}
\begin{align}\label{alphaGY}
\begin{split}
    \alpha_g\big|_{\texttt{311}} &= 
		\frac{3}{2\sqrt{7}} \frac{\nc}{\sqrt{\nc^2 -\frac{33}{112}}} \frac{1}{\sqrt{\epsilon}} 
		+\frac{9}{28} \frac{\nc^2 +\frac{3}{2}}{\nc^2 -\frac{33}{112}} \frac{1}{\epsilon} 
        + O\left(\frac{1}{\epsilon^{3/2}}\right)
\\
    \alpha_y\big|_{\texttt{311}} &= 
		\frac{9}{2\sqrt{7}} \frac{\nc^2 -1}{\nc\sqrt{\nc^2 -\frac{33}{112}}} \frac{1}{\epsilon^{3/2}} 
		+\frac{27}{28} \frac{\left(\nc^2 -1\right)\left(\nc^2 +\frac{3}{2}\right)}{\nc^2 \left(\nc^2 -\frac{33}{112}\right)} \frac{1}{\epsilon^2} +O\left(\frac{1}{\epsilon^{5/2}}\right)
\\
   \alpha_u\big|_{\texttt{311}} &= 
		\frac{9}{4\sqrt{7}} \frac{\nc^2 -1}{\nc\sqrt{\nc^2 -\frac{33}{112}}} \frac{1}{\epsilon} 
		+\frac{27}{56} \frac{\nc^2 -1}{\nc} 
\left(\frac{\nc^2 +\frac{3}{2}}{\nc^2 -\frac{33}{112}} -\frac{\sqrt{7}}{3}\frac{\nc}{\sqrt{\nc^2 -\frac{33}{112}}}  \right)\frac{1}{\epsilon^{3/2}} 
		+O\left(\frac{1}{\epsilon^{2}}\right)
\\
    \alpha_v\big|_{\texttt{311}} &= 
		-\frac{9}{4\sqrt{7}} \frac{\nc^2 -1}{\nc\sqrt{\nc^2 -\frac{33}{112}}} \frac{1}{\epsilon} 
				-\frac{27}{56} \frac{\nc^2 -1}{\nc^2}  
		\left(\frac{\nc^2 -\frac{3}{2}}{\nc^2 -\frac{33}{112}} -\sqrt{7}\frac{\nc}{\sqrt{\nc^2 -\frac{33}{112}}}\right)\frac{1}{\epsilon^{3/2}} 
		+O\left(\frac{1}{\epsilon^{2}}\right)
\end{split}
\end{align}
\end{widetext}
Notice that the gauge coupling achieves the exact same fixed point as in \eq{alphaBZ} to leading order in $\eps$ and $\nc$, 
\beq
\alpha_g^*|_{\rm BZ}=\alpha_g^*|_{\rm GY}\,,
\eeq
modulo subleading terms. Moreover, to the leading order in $\eps\gg 1$, the Yukawa and quartic couplings are suppressed as 
\begin{align}
\begin{split}
\alpha_y^*&\sim \alpha^*_g/(\nc^2\,\eps)\ll \alpha_g^*\\
\alpha_{u,v}^*&\sim \alpha_g^*/(\nc^2\sqrt{\eps})\ll \alpha_g^*
\end{split}
\end{align}
compared to the gauge coupling. In this light, the putative gauge-Yukawa fixed point \eq{alphaGY} where $\alpha^*_{y,u,v}|_{\rm GY}\to 0$ 
 for $1/\eps\to 0$, becomes degenerate with
the putative Banks-Zaks fixed point \eq{alphaBZ}  where $\alpha^*_{y,u,v}|_{\rm BZ}=0$. The degeneracy is only lifted through subleading corrections in $1/\eps$,
which leave a mild effect on the lower bound for the conformal window. The result for the putative large-$\eps$ fixed points is displayed 
by the two red lines in Fig.~\ref{fig:cw_sl},  and in comparison with the small-$\eps$ results at weak coupling (blue line).

Incidentally, the scalar couplings in  \eq{alphaGY} take identical values with opposite sign at the leading order in $1/\eps$,  
implying a flat potential.
In order to assess vacuum stability we must go to the next-to-leading order, which is found to generate
an overall positive sign for the potential,
\begin{align}
U^*\big|_{\texttt{311}} &= \frac{9}{4\sqrt{7}} \frac{ \nc^2-1 }{\nc \sqrt{\nc^2 -\frac{33}{112}}} \frac{1}{\epsilon^{3/2}}\,.
\end{align}
We conclude from the explicit expressions that vacuum stability does not  constrain the gauge Yukawa fixed point. 

Turning to the conformal window of the fixed point  \eq{alphaGY}, we find that it is bounded
 from below  by strong coupling. Following the same reasoning
as in the previous sections,
the boundary $\epsilon_{\rm subl.}$ is given by the largest positive root of the fourth-order polynomial,
\begin{align}
\begin{split}
    f(x) &=
		x^4\left(1792 \nc^4-528 \nc^2\right) \\
&\mathrel{\phantom{=}}
		+x^3\left(21496 \nc^4-4296 \nc^2-9504\right) \\
&\mathrel{\phantom{=}}
		+x^2\left(26140 \nc^4+41172 \nc^2-168048\right) \\
&\mathrel{\phantom{=}}
		+x\left(-406566 \nc^4+495858 \nc^2-985824\right) \\
&\mathrel{\phantom{=}}
		-1044531 \nc^4+1250172 \nc^2-1919808
\end{split}
\end{align}
The solution  turns out to be a smooth function of $\nc$ and can be approximated by the interpolant
\begin{align}
    \epsilon_{\rm subl. min} &=  \frac{4.2922\left(\nc^2 -0.6408\right)^2}{\left(\nc^2 -1.550\right)\left(\nc^2 +0.8046\right)}
\end{align}
from which we can directly read off the asymptotic value in the Veneziano limit. 
In Fig.~\ref{fig:cw_sl} we additionally compare the conformal windows at small and large $\eps$ using $\nc=7$.
For small $\eps$, couplings and scaling exponents at the gauge-Yukawa fixed point remain perturbative throughout, for all approximations up to \texttt{322}. For large $\eps$, couplings become very large close to the lower bound  for the putative Banks-Zaks (\texttt{300} approximation) and the putative gauge-Yukawa  (\texttt{311} approximation) fixed points. The small difference between the latter two conformal windows is due to subleading terms in $\eps$ due to Yukawa interactions.

It is now straightforward to determine the eigenvalues of the stability matrix following the same procedure as before. An interesting feature is 	that we find one relevant eigenvalue that scales as $\vartheta_1 \sim \sqrt{\epsilon}$. It is intriguing that although the couplings scale with negative powers of $\epsilon$, this eigenvalue does not become parametrically smaller, but larger, in the large $\epsilon$ regime. Large eigenvalues tend to be associated to non-perturbative phenomena, and these results from perturbation theory cannot be viewed as reliable. 
To next-to-leading order, the scaling exponents are
\begin{widetext}
\begin{align}\label{thetaGY}
    \begin{split}
    \vartheta_1\big|_{\texttt{311}} &= 
		-\frac{4}{\sqrt{7}}\frac{\nc}{\sqrt{\nc^2 -\frac{33}{112}}}\sqrt{\epsilon} 
		+\frac{3}{14}\left(\frac{\nc^2 -21}{\nc^2 -\frac{33}{112}}\right) 
 		+\frac{117}{6\sqrt{7}} \left(\frac{\nc^4 +\frac{814}{273}\nc^2 +\frac{369}{273}}{\nc\left(\nc^2 -\frac{33}{112}\right)^{3/2}}\right) \frac{1}{\sqrt{\epsilon}} \\
    &\mathrel{\phantom{=}}
    -\frac{32589}{1568} \left(\frac{\nc^6 -\frac{14701}{4828}\nc^4 -\frac{3552}{4828}\nc^2 -\frac{4347}{4828}}{\nc^2\left(\nc^2 -\frac{33}{112}\right)^2}\right) \frac{1}{\epsilon} 
		+O\left(\frac{1}{\epsilon^{3/2}}\right)
\\
    \vartheta_2\big|_{\texttt{311}} &= 
		\frac{9}{\sqrt{7}}\frac{\nc^2 -1}{\nc\sqrt{\nc^2 -\frac{33}{112}}}\frac{1}{\sqrt{\epsilon}}
		+\frac{351}{56}\frac{(\nc-1)(\nc+1)\left(\nc^2-\frac{3}{13}\right)}{\nc^2\left(\nc^2 -\frac{33}{112}\right)}\frac{1}{\epsilon} 
		+O\left(\frac{1}{\epsilon^{3/2}}\right)
\\
    \vartheta_3\big|_{\texttt{311}} &= 
		\frac{36}{\sqrt{7}}\frac{\nc^2 -1}{\nc\sqrt{\nc^2 -\frac{33}{112}}}\frac{1}{\epsilon}  
		+\frac{54}{7}\frac{(\nc-1)(\nc+1)\left(\nc^2 +\frac{3}{2}\right)}{\nc^2\left(\nc^2 -\frac{33}{112}\right)}\frac{1}{\epsilon^{3/2}} 
		+O\left(\frac{1}{\epsilon^{2}}\right)
\\
    \vartheta_4\big|_{\texttt{311}} &= 
		\frac{18}{\sqrt7}\frac{\nc^2-1}{\nc\sqrt{\nc^2 -\frac{33}{112}}}\frac{1}{\epsilon} 
		+\frac{27}{\sqrt{7}}\frac{\left(\nc^2-1\right)\left(\frac{2\nc^2+3}{2\sqrt{7}} +\nc\sqrt{\nc^2 -\frac{33}{112}}\right)}{\nc^2\left(\nc^2 -\frac{33}{112}\right)}\frac{1}{\epsilon^{3/2}} 
		+O\left(\frac{1}{\epsilon^{2}}\right)
\end{split}
\end{align}
\end{widetext}
The result confirms that the  relevant scaling exponent grows in the same way as it would at the putative Banks-Zaks fixed point
to leading order in $\eps\gg 1$,
\beq
\vartheta_1|_{\rm BZ}=\vartheta_1|_{\rm GY}\,,
\eeq
modulo subleading corrections. All other exponents become 
parametrically small  in the \texttt{311} approximation, and vanish asymptotically   in the limit of large $\eps$,
\beq
\vartheta_{2,3,4}|_{\rm GY}\to 0\,.
\eeq 
As such, the scaling dimensions are  equivalent to those of the Banks-Zaks fixed point at the leading order in $1/\eps$. 

\begin{figure}[b]
\includegraphics[width=.85\linewidth]{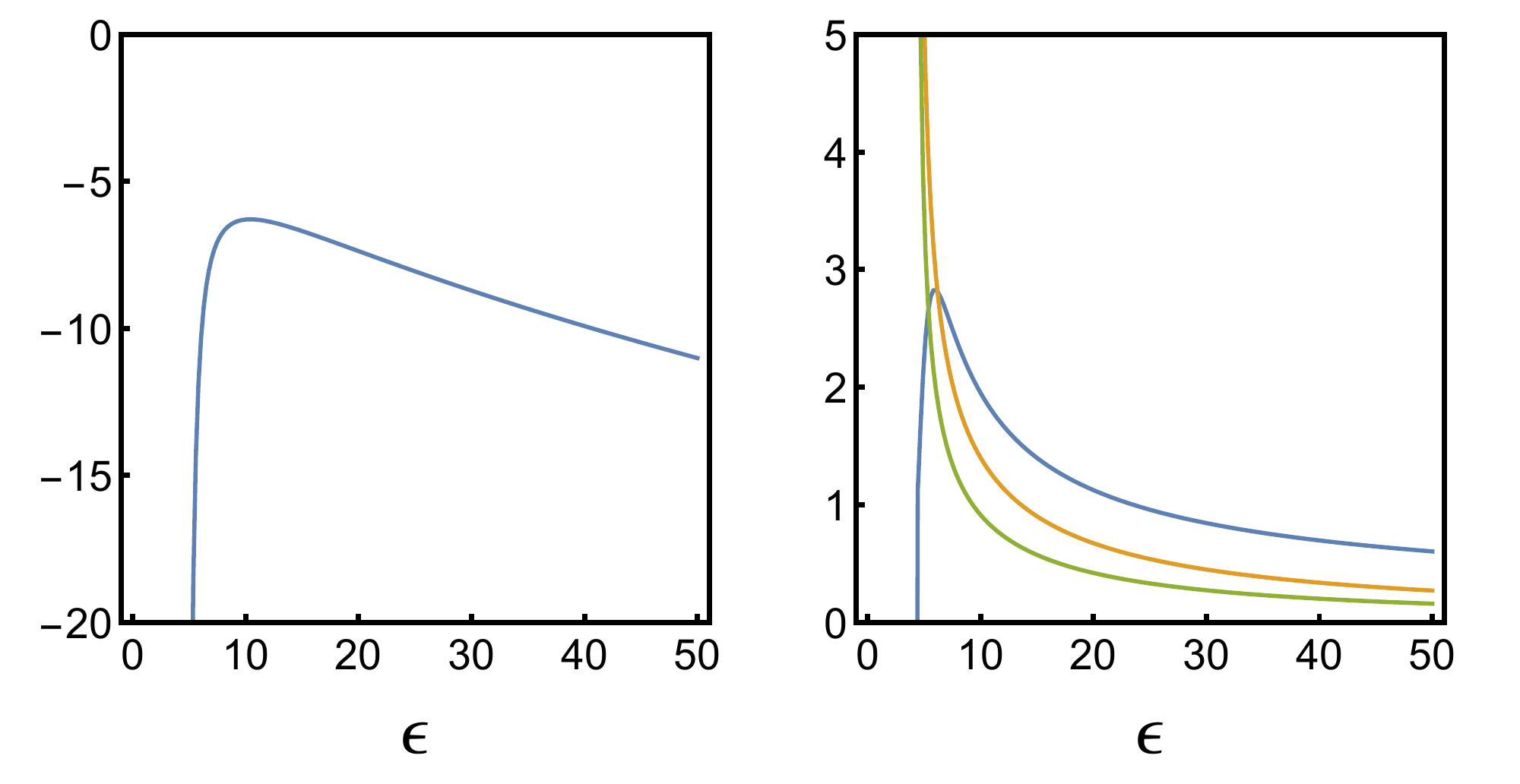}
\caption{Shown are the eigenvalues of the  gauge-Yukawa fixed point in the large $\epsilon$ regime ({\texttt{311}} approximation, $\nc=5$). The sole relevant scaling exponent comes out large $|\vartheta|>5$ (left panel). Irrelevant exponents may be similarly large close to the endpoint, and vanish for asymptotically large $\eps$.}
\label{fig:fp3_ev}
\end{figure}

We can further probe the region close to the lower boundary of the conformal window, where we find that eigenvalues grow large as they approach the boundary. This is illustrated in Fig.~\ref{fig:fp3_ev}, where the left panel shows the relevant eigenvalue 	diverging at both $\epsilon \to \infty$ and $\epsilon \to \epsilon_{\rm subl.min}$, and the right panel shows the irrelevant eigenvalues as a function of $\epsilon$ for $\nc=5$. 

The next natural step is to continue pushing this further to include the two loop running of the Yukawa and scalar interactions. For now, we include the two loop term of the Yukawa coupling only, leaving the scalars at one loop. We will see that the scaling does not work appropriately any more, but to demonstrate this a little work is needed. For simplicity we take the limit $\nc \to \infty$, but our results hold for finite $\nc$ as well. 

Keeping the self-interaction running to one loop means we can continue using the scaling $\alpha_{(u,v)} \sim \sqrt{\epsilon}\, \alpha_y$. Substituting this into the two loop running of the Yukawa beta function, we notice that actually the scalar contributions are subleading for large $\epsilon$ and can be neglected at leading order
\begin{align}
\beta_y^{(2)}\bigr|_{\alpha_{(u,v)} \propto \sqrt{\epsilon}\, \alpha_y} &\approx \frac{10}{3}\epsilon\, \alpha_g^2 \alpha_y + 8 \epsilon\, \alpha_g \alpha_y^2 -\frac{\epsilon^2}{2} \alpha_y^3 
\end{align}
To proceed we could assume the same scaling for $\alpha_g$ and $\alpha_y$ as before, however we would run into trouble as then the term that is proportional to $\epsilon\, \alpha_g^2 \alpha_y$ would become more relevant than the one loop terms, i.e. the Yukawa beta function does not vanish,
\begin{align}
\begin{split}
	\beta_y^{(1)}\bigr|_{\alpha_y \propto \alpha_g/\epsilon} &\sim \frac{\alpha_g^2}{\epsilon} \sim \frac{1}{\epsilon^2}\\
	\beta_y^{(2)}\bigr|_{\alpha_y \propto \alpha_g/\epsilon} &\sim \alpha_g^3 \sim \frac{1}{\epsilon^{3/2}}\,,
\end{split}
\end{align}
which suggests that we should use a different scaling. We allow $\alpha_y$ to change and keep the scaling of $\alpha_g$ as it is. This is indeed what we would expect as the scaling of the gauge coupling should only be determined by the fermion contributions in the gauge beta function. The only way to cancel the $\epsilon\, \alpha_g^2 \alpha_y$ term is by choosing $\alpha_y \propto \alpha_g/\sqrt{\epsilon} \sim 1/\epsilon$, so that
\begin{align}
\begin{split}
	\beta_y^{(1)}\bigr|_{\alpha_y \propto \alpha_g/\sqrt{\epsilon}} &\sim \alpha_g^2 \sim \frac{1}{\epsilon}\\
	\beta_y^{(2)}\bigr|_{\alpha_y \propto \alpha_g/\sqrt{\epsilon}} &\sim \sqrt{\epsilon}\, \alpha_g^3 \sim \frac{1}{\epsilon}\,.
\end{split}
\end{align}
This leads to a cancellation between one and two loop terms, with the Yukawa interaction scaling as $\alpha_y^* = 2\sqrt{5/3}\,\alpha_g^*/\sqrt{\epsilon}$. However substituting this result in the gauge beta function we run into trouble again. The Yukawa contributions at three loop are now dominating over the fermion contributions, reverting the overall sign of the three loop term,
\begin{align}\label{noFP}
\beta_g^{(3)}\bigr|_{\alpha_y \propto \alpha_g/\sqrt{\epsilon}} &\sim \frac{-112 + 540}{27} \epsilon^2\, \alpha_g^4\,.
\end{align}
The cancellation between one and three loop orders can now only happen for unphysical values of the gauge coupling. It is easy to verify that changing the scaling of the gauge coupling does not help either, as this leads to terms that do not cancel in the large $\epsilon$ limit. Similarly, including the two loop running of the scalar self interactions does not resolve this issue.

\subsection{Discussion}
We have confirmed that a remnant of a putative ultraviolet Banks-Zaks fixed point at large $\nf$ is visible within  perturbation theory, in accord with  \cite{Pica:2010xq,Shrock:2013cca}.
In \cite{Alanne:2019vuk}, however, it has been explained  that this fixed point is incompatible within any finite set of higher-order corrections, which invalidates fixed point claims based on  singularities of large $\nf$ beta functions. We therefore conclude that the cancellation pattern leading to  \eq{alphaBZ} is, in fact, spurious, and that it does not extend into a reliable physical UV fixed point at higher orders.

We then have investigated whether the inclusion of uncharged scalar matter fields and Yukawa couplings may give rise to a qualitatively different cancellation patterns at large $\nf$, different from the one observed in the gauge sector alone. However, using the three loop gauge beta function together with the one loop scalar and Yukawa quartics, and to the leading order in large $\eps$, this leads to
 the same fixed point and scaling exponents as found previously, see \eq{alphaBZ} vs \eq{alphaGY} and \eq{thetaBZ} vs \eq{thetaGY}. Hence, the presence of Yukawa and quartic scalar couplings does not offer a new scaling limit, the gauge-Yukawa fixed point \eq{alphaGY} is structurally identical to the Banks-Zaks one and is expected, consequently, to suffer from the same shortcomings  \cite{Alanne:2019vuk}.
 
 The scaling of fixed point couplings with $\eps$ could have been different from the Banks-Zaks one, provided that two loop scalar and Yukawa corrections dominate. If this were the case, however, the overall  sign of the gauge beta function along the Yukawa nullcline remains positive and an interacting fixed point cannot be achieved in the gauge sector \eq{noFP}.
Therefore,  perturbation theory offers no indication for a qualitatively different  large $\nf$  cancellation pattern  due to  Yukawa interactions, different from the one observed with a gauge coupling alone. 
This is rather different from what has been observed at weak coupling, where Yukawa couplings open up phase space for fixed points different from Banks-Zaks ones.

In future work, more clarity could be provided through proper non-perturbative studies, or by
all-order resummations of models with gauge, Yukawa, and scalar interactions.\footnote{See \cite{Antipin:2018zdg, Kowalska:2017pkt} for examples of resummed Yukawa beta functions where only one pair of fermions couple to a complex scalar field.} For now, we conclude that  there is not   sufficient evidence for  the existence of Banks-Zaks or gauge-Yukawa fixed points in matter-dominated regimes of large $\eps$.

\section{\bf Discussion and Conclusions}\label{Conclusions}
We  have performed a comprehensive search for interacting fixed points and their conformal windows in  QED- and QCD-like theories characterised by the absence of asymptotic freedom. We have determined their asymptotically safe fixed points and the corresponding conformal windows at the highest available order in perturbation theory, and extended earlier results \cite{Bond:2017tbw} beyond the Veneziano limit. Most notably, we find 
that finite $N$ corrections consistently decrease the size of the  conformal window, albeit moderately  (see Figs.~\ref{fig:cw1},~\ref{fig:eps_max_all}, and~\ref{fig:cw2}).
  Another noteworthy feature  is the smallness of  the control parameter $\eps$ within the entire conformal window (Tab.~\ref{tbl:eps}) which ensures that fixed point interactions, even for moderate matter field multiplicities, remain   as perturbative as  QCD at   $Z$ pole mass energies  (Fig.~\ref{fig:alphaAS2}).

Increasing the number of fermion species, we established that the loss of conformality   arises due to a fixed point merger and the loss of vacuum stability. This implies the existence of a new conformal fixed point $\alpha^*_{\rm IR}$ which, here, takes the form of a fully attractive IR sink for all canonically marginal couplings (Fig.~\ref{fig:PD}). 
Close to the merger, the  fixed point is accessible in perturbation theory. Exactly at the merger, one of the eigenperturbations  becomes exactly marginal and Miransky scaling is observed. Beyond the merger,  both fixed points disappear into the complex plane 
and leave a regime of slowed-down RG evolution (``walking'') in their wake. 
As such, our models offer examples of $4d$  QFTs which asymptote towards interacting and unitary conformal field theories both in the deep UV and in the deep IR limits.

Our findings are also of interest in the context of QCD with $\nf$ flavours of fermions, where it has previously been speculated that a merger of the Banks-Zaks fixed point  with a putative new  conformal fixed point $\alpha^*_{\rm QCD}$  \cite{Kaplan:2009kr} is responsible for the lower bound of the conformal window.
Since 
the boundary arises at strong coupling, however, a clear confirmation of the conformal fixed point $\alpha^*_{\rm QCD}$ has thus far remained elusive. It would then seem promising to investigate the new IR fixed point $\alpha^*_{\rm IR}$ of this work more extensively, for it arises at weak coupling and may  serve as a well-controlled template for a genuine merger  in 4d quantum gauge theories.

Finally, we have also revisited putative  fixed points in the regime of large $\eps$  \cite{PalanquesMestre:1983zy, Gracey:1996he, Holdom:2010qs,Alanne:2019vuk}, where fluctuations are matter-dominated.  Our main result is that
gauge-Yukawa fixed points, if they arise,   do so through the same  mechanism as the putative large-$\nf$ Banks-Zaks fixed point. Alternative scaling 
relations failed to provide viable fixed points, and we  conclude that the addition of Yukawa and scalar couplings  do not offer a  fixed point candidate different from the  Banks-Zaks one. In this light, the objections brought forward against the large-$N_{\rm f}$ Banks Zaks fixed point    also apply for  the large-$N_{\rm f}$ gauge-Yukawa fixed point.  Hence, unlike earlier expectations,  resummed versions of perturbation theory  do not offer signatures for  viable fixed points in the matter-dominated large-$N_{\rm f}$ regime. 
It will  of course be important to further test these conclusions  beyond perturbation theory in the future.
\\[3ex]

{\bf Acknowledgements.} 
We thank Tom Steudtner for comments on the manuscript. 
DL is supported by the Science and Technology Facilities Council (STFC) under the Consolidated Grant ST/T00102X/1.
GMV has been supported by the {\it Consejo Nacional de Ciencia y Tecnologia} (CONACYT).

\appendix
\section{\bf Beta Functions}\label{Appendix}

We list here the beta function coefficients for the quantum field theory with Lagrangian \eq{L} with general compact simple gauge group $\gp$, $\nf$ Dirac fermions in an irreducible representation $R$
of it,   using known formal expressions provided in \cite{Machacek:1983tz, Machacek:1983fi, Machacek:1984zw, Luo:2002ti, Schienbein:2018fsw}.  
For a powerful tool to extract perturbative RG beta functions from general expressions in the $\overline{\rm MS}$ scheme, see \cite{Litim:2020jvl}.

Unlike in the main text  \eq{eq:alphasN}, we do not rescale the couplings  by powers of matter multiplicities, and write
\begin{equation}\label{eq:alphas}
\alpha_x=\frac{x^2}{\left(4\pi\right)^2}\,, \quad 	\alpha_u=\frac{u}{\left(4\pi\right)^2}		\,, \quad	\alpha_v=
\frac{v}{\left(4\pi\right)^2}
\end{equation}
where $x=g,y$. The reason for this is that the large $N$ scaling in the fundamental or other irreducible representations works quite differently.
We introduce the short-hand notation $\co{x}{\alpha_K\alpha_L\dots}$ to mean the coefficient of $\alpha_K\alpha_L\dots$ in $\beta_x$, i.e.
\begin{align}
	\co{x}{\alpha_K\alpha_L\dots} = \left.\frac{\partial \beta_x}{\partial \alpha_K\partial\alpha_L\dots}\right|_{\alpha = 0}\,.
\end{align}
We group the coefficients by coupling, and then by loop order.
Firstly, we have the gauge coupling, at one-loop,
\begin{align}
	\label{eq:beta_g1_R}
	\co{g}{\alpha_g^2} &= - \frac{4}{3}\left(\frac{11}{2}C_2^\gp - 2 \nf S_2^R\right)\,,
\end{align}
two-loop,
\begin{align}
	\label{eq:beta_g2_R}
	\co{g}{\alpha_g^3} &= 
		\left(8 C_2^R + \frac{40}{3} C_2^G\right) \nf S_2^R - \frac{68}{3} (C_2^G)^2\,, 	\\		
	\co{g}{\alpha_g^2\alpha_y} &= - 4 S_2^R \nf^2\,,
\end{align}
and three-loop,
\begin{align}
	\label{eq:beta_g3_R}
	\co{g}{\alpha_g^4} &= -6\left[\frac{2857}{162}(C_2^G)^3 - \frac{1415}{162} (C_2^G)^2(2\nf S_2^R)\right.\nonumber\\
	&+ \frac{79}{162} (C_2^G)(2\nf S_2^R)^2
	 - \frac{205}{54} C_2^G C_2^R(2\nf S_2^R) \nonumber\\
	 	&\left. + \frac{11}{27} C_2^R(2\nf S_2^R)^2
		+ \frac{1}{3} (C_2^R)^2(2\nf S_2^R)\right]\,,\\
	\co{g}{\alpha_g^3\alpha_y} &= -6\left[2 C_2^G (2\nf S_2^R) \nf 
			+ \frac{1}{2}C_2^R (2\nf S_2^R) \nf\right]\,,\\
	\co{g}{\alpha_g^2\alpha_y^2} &= 6\left[\frac{1}{2}\nf^2(2\nf S_2^R) 
			+\frac{7}{12}\nf d_R(2\nf S_2^R)\right]\,.
\end{align}
The Yukawa coupling coefficients at one-loop are
\begin{align}
	\co{y}{\alpha_y^2} &= 2(\nf + d_R)\,,	\\
	\co{y}{\alpha_y\alpha_g} &= 12 C_2^R\,,
\end{align}
and two-loop contributions from gauge and Yukawa
\begin{align}
	\co{y}{\alpha_y^3} &= 2\left[2 - \frac{1}{4} \nf^2 - 3 \nf d_R\right]\,,\\
	\co{y}{\alpha_y^2 \alpha_g} &= 2\left[ 8 \nf C_2^R+5 C_2^R d_R\right]\,,\\
	\co{y}{\alpha_y \alpha_g^2} &= 4\left[-\frac{3}{2}(C_2^R)^2 - \frac{97}{6}C_2^R C_2^G 
			+ \frac{5}{3}C_2^R(2\nf S_2^R)\right]\,,
\end{align}
as well as two-loop contributions involving the scalar quartics
\begin{align}			
	\co{y}{\alpha_y^2 \alpha_u} &= - 8 (\nf^2 + 1)\,,
	\\ \co{y}{\alpha_y^2 \alpha_v} &= - 8 \nf\,,\\
	\co{y}{\alpha_y\alpha_u^2} &= 4(\nf^2 + 1)\,,
	\\ 
	\co{y}{\alpha_y\alpha_v^2} &= 4(\nf^2 + 1)\,,\\
	\co{y}{\alpha_y\alpha_u \alpha_v} &=16 \nf \,.
\end{align}
Lastly, we have the scalar quartic beta functions at one-loop
\begin{align}
	\co{u}{\alpha_u^2} &= 8 \nf\,,		
	\\
	\co{u}{\alpha_u\alpha_v} &= 24\,,\\
	\co{u}{\alpha_u\alpha_y} &= 4 d_R\,,		\\
	\co{u}{\alpha_y^2} &= -2 d_R\,,\\
	\co{v}{\alpha_v^2} &= 4(\nf^2 + 4)\,,			
	\\
	\co{v}{\alpha_u\alpha_v} &= 16 \nf\,,\\
	\co{v}{\alpha_v\alpha_y} &= 4 d_R\,,		
	\\
	\co{v}{\alpha_u^2} &= 12\,,
\end{align}
and two-loop for the single-trace
\begin{align}
	\co{u}{\alpha_u^3} &= -24 (5 + \nf^2)\,,\\
	\co{u}{\alpha_u^2\alpha_y} &= -16 \nf d_R\,,\\
	\co{u}{\alpha_u^2\alpha_v} &= -352 \nf\,,\\
	\co{u}{\alpha_u\alpha_y^2} &= -6 \nf d_R\,,\\
	\co{u}{\alpha_u\alpha_v^2} &=-8(41+5 \nf^2) \,,\\
	\co{u}{\alpha_u\alpha_y\alpha_g} &= 20 C_2^R d_R\,,\\
	\co{u}{\alpha_u\alpha_y\alpha_v} &= - 48 d_R\,,\\
	\co{u}{\alpha_y^3} &= 4 \nf d_R\,,\\
	\co{u}{\alpha_g\alpha_y^2} &= -8 C_2^R d_R\,,\\
	\co{u}{\alpha_v\alpha_y^2} &= 8 d_R\,,
\end{align}
and for the double-trace
\begin{align}
	\co{v}{\alpha_v^3} &= -24 (7+3 \nf^2)\,,\\
	\co{v}{\alpha_v^2\alpha_y} &= -8d_R(4+\nf^2)\,,\\
	\co{v}{\alpha_v^2\alpha_u} &= -352 \nf\,,\\
	\co{v}{\alpha_v\alpha_y^2} &= -6 \nf d_R\,,\\
	\co{v}{\alpha_v\alpha_u^2} &= -8(41 + 5 \nf^2)\,,\\
	\co{v}{\alpha_v\alpha_y\alpha_g} &= 20 C_2^R d_R\,,\\
	\co{v}{\alpha_u\alpha_y\alpha_v} &= -32 \nf d_R \,,\\
	\co{v}{\alpha_u^3} &= -96 \nf\,,\\
	\co{v}{\alpha_u^2\alpha_y} &= - 24 d_R \,,\\
	\co{v}{\alpha_u \alpha_y^2} &= 8 d_R\,,\\
	\co{v}{\alpha_y^3} &= 4 d_R\,.
\end{align}
If we specialise to the case $\gp = SU(\nc)$ with the fermions in the fundamental representation $R = \nc$, we have
\begin{align}
	C_2^\gp &= \nc = d_R\,,	\\
	S_2^R &= \tfrac{1}{2}\,,	\\
	C_2^R &= \tfrac{1}{2}(\nc - \tfrac{1}{\nc})\,,
\end{align}
and we recover the equations given in the main text, once we have rescaled the couplings as in \eqref{eq:alphas} and rewritten in terms of $\epsilon$ \eqref{eq:eps}.

\bibliographystyle{apsrev4-1}
\bibliography{PriceASbib,refs_fun1}

\end{document}